\documentclass[aps,prr,floats,floatfix,superscriptaddress,nofootinbib,twocolumn,showkeys]{revtex4-2}
\usepackage[colorlinks=true, linkcolor=blue, citecolor=blue, urlcolor=blue]{hyperref}
\usepackage{bbm}
\usepackage{amsmath,amssymb,amsthm}
\usepackage{xcolor}
\usepackage{graphicx}
\usepackage{caption}
\usepackage{makecell}
\usepackage{url}
\usepackage{answers}
\usepackage{array}
\usepackage{multirow}
\usepackage[mathscr]{euscript}
\usepackage{eufrak}
\usepackage{calligra}

\usepackage{graphicx}
\usepackage{dcolumn}
\usepackage{bm}
\usepackage{booktabs}

\newcommand{\ZZZ}[2]{}

\newcommand{\quotes}[1]{``#1''}

\begin{document}

\preprint{APS/123-QED}

\title{\textbf{Geometric Organization and Inference of Shortest Path Nodes in Soft Random Geometric Graphs} 
}%

\author{Zhihao~Qiu}
\affiliation{
Faculty of Electrical Engineering, Mathematics and Computer Science, 
Delft University of Technology, 
2600 GA Delft, The Netherlands}
\author{S\'amuel~G.~Balogh}
\affiliation{
Faculty of Electrical Engineering, Mathematics and Computer Science, 
Delft University of Technology, 
2600 GA Delft, The Netherlands}
\affiliation{National Laboratory of Health Security, 
HUN-REN Alfréd Rényi Institute of Mathematics, 
1053 Budapest, Hungary}

\author{Xinhan~Liu}
\author{Piet~Van~Mieghem}
\author{Maksim~Kitsak}
\email{maksim.kitsak@gmail.com}
\affiliation{
Faculty of Electrical Engineering, Mathematics and Computer Science, 
Delft University of Technology, 
2600 GA Delft, The Netherlands}


\date{\today}

\begin{abstract}
The shortest path problem is related to many dynamic processes on networks, ranging from routing in communication networks to signaling in molecular interaction networks. When the network is fully known, the shortest path problem can be solved precisely and in polynomial time. If, however, the network of interest is only partially observable, the shortest path problem is no longer straightforward. Inspired by the shortest path problem in partially observable networks, we investigate the geometric properties of shortest paths in {\it Euclidean} Soft Random Geometric Graphs (SRGGs). We find that shortest paths are aligned along geodesic curves connecting shortest path endpoints.  The strength of the shortest path alignment, as quantified by the average distance to geodesic from shortest path nodes and the average path stretch, is higher for larger SRGGs with short-range connections. In addition, we find that the strength of the shortest path alignment is non-monotonic with respect to the average degree of the SRGG. Based on these observations, we establish the conditions under which the alignment of shortest paths may be sufficiently strong to allow the identification of shortest path nodes based on their proximity to geodesic curves. We show that in partially observable networks with uncertain node positions, our geometric approach can outperform network-based shortest-path algorithms. In practical settings, our findings may have applications to navigation, wireless routing, and flow characterization in infrastructure networks.

\end{abstract}

\keywords{shortest path, soft random geometric graph, distance to geodesic, geodesic curve, path finding}
\maketitle

\section{Introduction}\label{Sec:Introduction}
The shortest path $\mathcal{P}^*_{ij}$ from a node $i$ to a node $j$ in a network $G$ is an ordered sequence of distinct nodes to traverse from $i$ to $j$, such that the sum of the link weights along the path is minimized. The shortest path has been extensively studied in the literature due to its relevance to a large array of real problems in routing~\cite{wang1992analysis,magzhan2013review,ChangAgeneticalgorithm,bertsekas1982dynamic,johnson1977efficient,gomathi2018energy}, transportation~\cite{FU20063324,pallottino1998shortest,VANVLIET19787,zhang2024mapreduce,DU20133505}, spreading processes~\cite{kitsak2010identification,friedkin1991theoretical,moukarzel1999spreading,pei2013spreading,karunakaran2017spreading} and search~\cite{kitsak2023finding,chen2012reliable}. 

 In this work, we consider unweighted undirected networks. In unweighted networks, the minimization of the sum of the link weights reduces to the minimization of the number of links, i.e., the hopcount. Further, in undirected networks, shortest paths are symmetric: $\mathcal{P}^*_{ij} = \mathcal{P}^*_{ji}$. If the network of interest is fully observable, a large array of methods exist that allow solving the shortest path problem in polynomial time~\cite{cormen2022introduction,dijkstra1959note,bellman1958routing}. 
These \emph{conventional} methods are guaranteed to solve the shortest path problem precisely, provided the entire network is fully observable. This assumption is, unfortunately,  not met in many large networks.  Indeed, large-scale social, technological, and biological networks are often \emph{substantially} incomplete: the majority of links and/or nodes are not known~\cite{luck2020reference, ahmed2020survey,claffy2022challenges}.

The accuracy of conventional methods quickly decreases as the number of missing network links increases~\cite{kitsak2023finding}, due to the extreme sensitivity of shortest paths: even a single missing (false negative) or a spurious (false positive) link may change the path entirely. The conventional shortest-path methods, including the popular Dijkstra~\cite{dijkstra1959note,cormen2022introduction} and Bellman-Ford algorithms~\cite{bellman1958routing,schrijver2012history,cormen2022introduction,ford1956network} rely on the iterative exploration of network topology and are very likely to fail if some of the links or nodes are not known~\cite{kitsak2023finding}.

A recent work proposed a novel method for finding shortest and nearly shortest path nodes, making use of their geometric alignment in hyperbolic embeddings of these networks: shortest path nodes are likely found in the geometric vicinity of geodesic curves connecting shortest path end points~\cite{kitsak2023finding}. To find a shortest path between nodes $A$ and $B$ in network $G$, one needs to obtain a hyperbolic representation of the network first, i.e. map network nodes to points in a hyperbolic disk. Node coordinates obtained as a result of hyperbolic embeddings allow one to draw a geodesic curve $\gamma(A,B)$ connecting shortest path endpoints $A$ and $B$. The likelihood for a node $C$ to belong to a shortest path between $A$ and $B$ can be estimated using the distance to geodesic $d\left(C, \gamma(A,B)\right)$ from  $C$ to $\gamma(A,B)$, the smaller the distance the higher the likelihood.

Kitsak et al, Ref.~\cite{kitsak2023finding} have demonstrated that hyperbolic embeddings can be used to find shortest paths even in extremely incomplete networks, with as few as $10\%$ of network links present, leading to promising applications in biomedicine and communication networks. The caveat of the method is that it is limited to networks whose organization is consistent with hyperbolic geometry. It is generally accepted that networks with scale-free degree distributions and strong clustering coefficients are best embedded into hyperbolic spaces~\cite{boguna_network_2021}.

In our work, we ask if shortest paths in {\it Euclidean} geometric networks are sufficiently aligned and identifiable with geometric methods. On the one hand, our work is fueled by earlier findings in {\it Euclidean} random geometric graphs demonstrating a certain degree of alignment of shortest paths~\cite{Kartun-GilesShape,diaz2016relation,ellis2007random,bradonjic2010efficient,friedrich2013diameter}. On the other hand, our work is inspired by the observation that many networks of interest are embedded into {\it Euclidean} spaces. 
Indeed, nodes of wireless communication networks are base stations placed in various geographic areas, and the quality of communication between any two base stations depends on the {\it Euclidean} geographic distance between them~\cite{VossDistributional,Raftopouloudronenetworks2022, hekmat2006connectivity,ZhangConnectivity}. Another example is transportation networks, where nodes are intersections and road segments connecting them are links~\cite{boyaci2021vehicle,buczkowska2019comparison}. 
Finally, of relevance to this study are critical infrastructure networks, e.g., power grids consisting of power stations connected with transmission lines~\cite{app8050772,IslambekovROLE}. In all these networks, nodes are characterized by their geographic positions, and links are typically connecting nearby nodes.

We conduct our studies in the context of the Soft Random Geometric Graph~\cite{PenroseSRGG, Kartun-GilesShape, mao2012connectivity}, which can be regarded as a special case of the 
Geometric Inhomogeneous Random Graph (GIRG) model~\cite{bringmann2019geometric, blasius2022efficiently}. In an SRGG,
nodes are positioned at random in a latent metric space $\mathcal{M}$, and connections between nodes are established independently with probabilities that are functions of distances between nodes in $\mathcal{M}$. We find that in SRGGs, shortest path nodes tend to be significantly close to geodesic curves connecting path endpoints than expected by chance. In other words, we establish that shortest paths are geometrically aligned along geodesic curves. These alignments exhibit non-trivial behavior as a function of graph properties such as the average degree and the clustering coefficient. We find that under optimal conditions, the alignment of shortest paths allows one to find them using node coordinates in the latent space. Further, we demonstrate that the geometric path-finding methods have a competitive advantage over conventional path-finding methods in situations when network topology is not completely observable.

The remainder of this paper is organized as follows. In Sec.~\ref{Sec:Methods}, we formalize the problem and introduce the Soft Random Geometric Graph (SRGG) model, together with the simulation framework used to quantify shortest-path alignment and to infer shortest-path nodes based on their distance to geodesic curves. In Section~\ref{sec:Quantifying the alignment of shortest paths along geodesic curves in SRGG models}, we study the alignment of shortest paths in Soft Random Geometric Graphs as a function of their parameters. Relying on the results of Section~\ref{sec:Quantifying the alignment of shortest paths along geodesic curves in SRGG models}, we examine the problem of the identification of shortest path nodes in Section~\ref{sec:Identifying shortest path nodes in SRGG with noise}. We conclude our work with the discussion and outlook in Section~\ref{sec:Conclusion}.

\section{Methods}\label{Sec:Methods}
\subsection{Random Geometric Graph (RGG) and Soft Random Geometric Graph (SRGG)}\label{sec:RGG}
A random geometric graph (RGG) ~\cite{dall2002random,penrose2003random} is an undirected graph constructed by randomly placing $N$ nodes into a metric space $\mathcal{M}$ and connecting any two nodes $i$ and $j$ if the distance $d_{ij}$ between them is less than a certain radius $r$. 
In our work, we consider RGGs built in a 2-dimensional unit square, where $N$ nodes are placed uniformly at random. If the  connection radius $r\ll 1$, the link density in the RGG with $L$ links, 
\begin{equation}
\label{eq:rggdensity}
p_{\rm RGG} \equiv \frac{L}{L_{\rm max}} \approx \pi r^{2},
\end{equation}
where $L_{\rm max} = \frac{1}{2} N(N-1)$, see, e.g.,~\cite{van2001paths}.

Thus, ignoring boundary effects, the probability that a node $i$ has exactly $k$ connections is 
\begin{equation}
\textrm{Pr}[D_i=k] = {N-1 \choose k}\left(p_{\rm RGG}\right)^k \left(1-p_{\rm RGG}\right)^{N-k-1},
\end{equation}
and the expected degree of a node is
\begin{equation}
    \mathbb{E}[D_i] = (N-1) p_{\rm RGG},
    \label{eq:ED_rgg}
\end{equation}
where $p_{\rm RGG}$ is the link density given by Eq.~(\ref{eq:rggdensity}).

A soft random geometric graph (SRGG) can be regarded as a generalization of the random geometric graph in which links are established independently with probabilities that are functions of distances between the nodes in $\mathcal{M}$. An SRGG can be generated as follows: 
\begin{enumerate}
    \item Sprinkle $N$ nodes with probability density function (pdf) $\rho(x)$ into latent space $\mathcal{M}$.
    \item Calculate the distances $d_{\mathcal{M}}(ij)$ between each $(i,j)$ node pair.
    \item Connect each $(i,j)$ node pair independently with probability $p_{ij}$, which is a function of  distances $d_{\mathcal{M}}(ij)$ between nodes $i$ and $j$, $p_{ij} = f\left(d_{\mathcal{M}}(ij)\right)$.
\end{enumerate}

While, in principle, $f:R^+\rightarrow [0,1]$ can be any function, $f$ is often selected to be a decreasing function to model networks where connections over short distances are preferable. One common choice for $f$ is 
\begin{equation}
     f(d) = \beta e^{- \left(\frac{d}{d_0}\right)^{\eta}},
     \label{eq:waxman_conn}
\end{equation}
where $\beta \in (0,1]$, $d_0 > 0$, and $\eta > 0$ are tunable parameters. In particular, the $\eta =1$ case is known as the Waxman graph model~\cite{penrose2003random,waxman2002routing}, and $\eta= \infty$ case reduces the SRGG model to the RGG. The choice of $\eta \in [2,6.5]$ corresponds to the Rayleigh fading connection function~\cite{ErbaRandom,Raftopouloudronenetworks2022} used to  model radio signal's amplitude and phase fluctuations in multipath environments.

In this work, we study SRGGs with a different connection probability function, which is inspired by the Fermi-Dirac statistics: 
\begin{equation}
     f(d) = \frac{1}{1+\left(\frac{d}{d_0}\right)^\beta},
     \label{eq:decreasing function for SRGG}
\end{equation}
where $d_0 >0$ and $\beta > 0$ are again parameters tuning topological properties of resulting graphs. The SRGG model is closely related to the $\mathbb{S}^{D}$ heterogeneous graph model~\cite{Krioukov2010hyperbolic} if node hidden variables in $\mathbb{S}^{D}$ are constant, and is also closely related to the Geometric Inhomogeneous Random Graph (GIRG) model~\cite{Boguna2019small}.

An SRGG model with assigned node coordinates can be regarded as a system with quantum states with energy levels $\epsilon_{ij}$ are functions of distances $d_{ij}$ between nodes~\cite{Krioukov2010hyperbolic}. Links in the SRGG can, therefore, be regarded as fermions occupying corresponding quantum states, at most one fermion per state~\cite{Krioukov2010hyperbolic}. Parameter $\beta$ then corresponds to the inverse temperature $\beta = \frac{1}{\tau}$. At low temperatures (high $\beta$ values) only quantum states with the lowest energy levels are occupied, i.e., only short-distance connections are possible. Higher temperatures (low $\beta$ values), on the other hand, allow the occupation of higher energy states, which correspond to longer distance connections in the SRGG. Following this analogy, we refer to $\beta$ as the inverse temperature parameter throughout the text.

\begin{figure}[!]
    \centering
    \includegraphics[width = 0.48\textwidth]{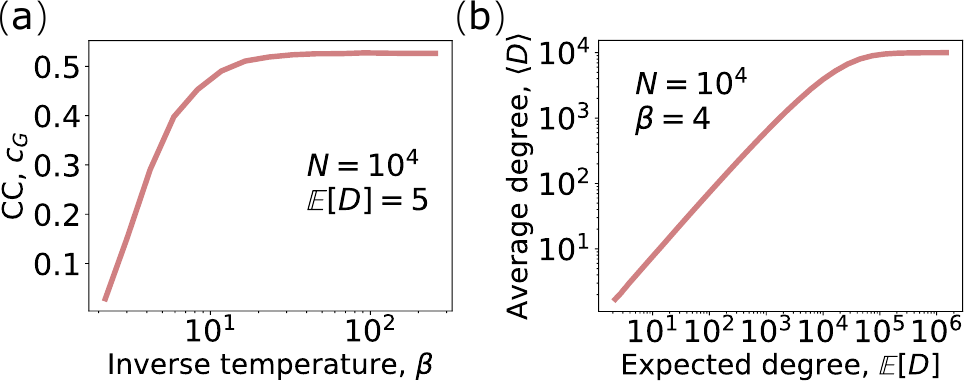}
    \caption{ \footnotesize (a) shows an example of how the clustering coefficient changes with the input inverse temperature parameter in SRGGs. (b) shows an example of how the real average degree changes with the input expected degree in SRGGs.
    The x-axis is the input expected degree, while the y-axis is the real average degree.}
    \label{fig:SRGG property}
\end{figure}

We consider the case of SRGGs built on a unit square, with uniform node density, $\rho(\mathbf{x}) = 1$. Ignoring the boundary effects, one can establish that the expected degree of the SRGG model is given by
\begin{equation}
\mathbb{E}[D] = \mathbb{E}\left[D_i\right] \approx \frac{cNd^2_0}{\beta \sin{\frac{2\pi}{\beta}}},
\label{eq:exp_degree_srgg}
\end{equation}
when $\beta > 2$, and the proportionality coefficient $c>0$, arizing due to finite size effects, can be tuned numerically.  Eq.~(\ref{eq:exp_degree_srgg}) indicates that parameter $d_0$ can be used to tune the expected degree of SRGG graphs. Indeed, greater $d_0$ values effectively rescale distances in the SRGG, allowing nodes to connect to other nodes over larger distances.  Eq.~(\ref{eq:exp_degree_srgg}) holds well for small and moderate expected degree values, Fig.~\ref{fig:SRGG property}(b). For large expected degree values, the boundaries of the unit square become non-negligible, making Eq.~(\ref{eq:exp_degree_srgg}) less precise. As a result, in the paper we carefully distinguish the theoretical expected degree $\mathbb{E}[D]$ of the SRGG model, from the realized average degree $\langle D\rangle$ of the SRGG, which we measure as $\langle D \rangle \equiv 2L/N$.

While inverse temperature $\beta$ also affects the average degree, it plays a far more important role by tuning the effective range of connections. When $\beta$ is large connections are mostly limited to distances $d \leq d_0$. Smaller $\beta$ values increase the probability of connections at larger distances. As a result, $\beta$ allows one to tune the average clustering coefficient in SRGGs, see Fig.~\ref{fig:SRGG property}(a). 

In our work, we parametrize SRGGs by the inverse temperature parameter $\beta$ and expected degree $\mathbb{E}[D]$. To generate an SRGG with desired parameters $\beta$ and $\mathbb{E}[D]$, we use Eq.~(\ref{eq:exp_degree_srgg}) to determine parameter $d_0$. We then use parameters $d_0$ and $\beta$ to determine connection probabilities for all node pairs in the network.

\subsection{Alignment of Shortest Paths Along Geodesic Curves}
\label{sec:methods:alignment}

A geodesic $\gamma(i,j)$ is the shortest length curve connecting points $i$ and $j$ in space $\mathcal{M}$. The length of the geodesic $\gamma(i,j)$ is the distance $d_{ij}$ between points $i$ and $j$. In the case of \emph{Euclidean} space, the geodesic $\gamma(i,j)$ is a straight line between $i$ and $j$, while the distance $d_{ij}$ represents the corresponding \emph{Euclidean} distance.

The distance $d(q,\gamma(i,j))$ from a point $q$ to the geodesic $\gamma(i,j)$ is the distance from point $q$ to point $x \in \gamma(i,j)$ on the geodesic, such that distance $d(q,x)$ is minimized. 
\begin{eqnarray}
d(q,\gamma(i,j)) &\equiv& {\rm min}~d(q,x),\\
&{\rm s.t.}&~ x \in \gamma(i,j).
\end{eqnarray}

In \emph{Euclidean} spaces, the distance from the point to a geodesic $d(q,\gamma(i,j))$ is the length of the perpendicular line segment from $q$ to the geodesic line $\gamma(i,j)$.

The hopcount $h_{ij}$ of a path from a node $i$ to a node $j$ is 
the number of links of that path from node $i$ to node $j$. Further, we define the path stretch $S_{ij}$ as the total geometric distance accumulated along the path $\mathcal{P}^*_{ij}$: 
\begin{equation}
S_{ij} \equiv \sum_{l \in \mathcal{P}^*_{ij}} d_l    
\end{equation}
where $d_l$ is the {\it Euclidean} distance between the two end nodes of link $l$. Fig.~\ref{fig:deviation}~(a) depicts an example of a shortest path, a geodesic, the distance from a node on the path to the geodesic, and the shortest path stretch in an SRGG.

In an unweighted SRGG, there can be more than one shortest path $\mathcal{P}_{ij}$ between a node pair $(i,j)$. The set $W_{P^{*}_{ij}}$ of all nodes constituting shortest paths between $i$ and $j$ are called \quotes{shortest path nodes}, and $|W_{P^{*}_{ij}}|$ is the cardinality, i.e., the number of elements in the set.

We quantify the alignment of shortest paths along geodesic curves with two methods. The first method is to compare distances from shortest path nodes to the geodesic with those from randomly selected nodes. The second method is to compute the shortest path stretch, i.e., the distance accumulated along shortest path links. The smaller the stretch the stronger the alignment, Fig.~\ref{fig:deviation}(a).

To assess the extent of the geometric alignment of shortest path nodes in SRGGs with given parameters, we design the following experiments: 
\begin{enumerate}
    \item Generate an SRGG with the network size $N$, the expected degree $\mathbb{E}[D]$ and the inverse temperature parameter $\beta$. We then select a number of node pairs at random. 
    \item For each node pair $(i,j)$, we obtain all shortest path nodes $W_{\mathcal{P}^*_{ij}}$.
    \item Find the geodesic curve $\gamma_{ij}$ connecting node $i$ and $j$.
    
    \item Compute the distance to geodesic $d(q, \gamma(i,j))$ for all shortest path nodes $q \in W_{\mathcal{P}^*_{ij}}$ except for node $i$ and $j$. Record the average, maximum and minimum distance for all shortest path nodes.
    
    \item Randomly select $m \equiv | W_{\mathcal{P}^*_{ij}}|-2$ nodes from the graph and compute the distance to the geodesic for each node.
    
    \item If there is a single shortest path connecting $i$ and $j$, compute its stretch. If there are multiple shortest paths connecting $i$ and $j$, select one shortest path at random and compute its stretch.
    
\end{enumerate}

\subsection{Predicting Shortest Path Nodes Using Distance to Geodesic}
\label{sec:method predict}
Kitsak et al.~\cite{kitsak2023finding} proposed a  method to infer shortest path nodes by measuring node proximities to geodesic curves connecting shortest path endpoints in hyperbolic embeddings of networks. 
We extend this approach to {\it Euclidean} SRGGs. We refer to this method as the \quotes{distance to the geodesic}.  

To infer shortest path nodes connecting nodes $i$ and $j$ of an SRGG in latent space $\mathcal{M}$, we first find the geodesic $\gamma(i,j)$ connecting the nodes in $\mathcal{M}$. Then, for each network node $q$ excluding $i$ and $j$, we find the distance to the geodesic $d\left(q, \gamma(i,j)\right)$, and rank all network nodes in increasing order of distance to the geodesic. The closest to the geodesic nodes are then viewed as likely candidates for the shortest path nodes.

As a reference for comparison, we also infer shortest path nodes by first reconstructing network links and then identifying shortest path nodes using the Dijkstra algorithm.
We consider two reconstruction strategies, denoted as ${\rm SRGG}+{\rm net}$ and ${\rm RGG}+{\rm net}$.

In the ${\rm SRGG}+{\rm net}$ experiment, we use known parameters $\beta$ and $\mathbb{E}[D]$ and node coordinates $\{\mathbf{x}_{i}\}$ and $\{\mathbf{y}_{i}\}$ to rebuild $10^{3}$ random graph instances $G_{\ell}$, $\ell = 1,...,10^{3}$. 
For each instance $\ell$, we find the set of shortest path nodes $W_{\mathcal{P}^{*}_{ij}(\ell)}$ using the Dijkstra algorithm on graph $G_{\ell}$. 
As a result, shortest path node candidates are nodes that appear most frequently in all $W_{\mathcal{P}^{*}_{ij}(\ell)}$ sets. 

The ${\rm RGG}+{\rm net}$ experiment is a special case of the ${\rm SRGG}+{\rm net}$ experiment when $\beta=\infty$. 
Since $\beta=\infty$ case corresponds to connecting nodes to geometrically closest nodes, the network reconstruction procedure is equivalent to the reconstruction of random geometric graphs, explaining the name of the experiment.
For a specific network size $N$, expected degree $\mathbb{E}[D]$ and node coordinates $\{\mathbf{x}_{i}\}$ and $\{\mathbf{y}_{i}\}$, there exists a unique RGG realization.
Hence, the RGG experiment reconstructs only one graph $G$.
All shortest path nodes $W_{\mathcal{P}^{*}_{ij}}$ in graph $G$ are predicted as shortest path node candidates by our RGG strategy.

To evaluate the accuracy of our geometry-based method, we use the statistical precision metric~\cite{powers2020evaluation,vujovic2021classification}. Our evaluation procedure can be summarized as follows:
\begin{enumerate}
    \item Generate an SRGG $G$ with required parameters $N$, $\mathbb{E}[D]$ and $\beta$. Select a number of node pairs in $G$ uniformly at random.
    
    \item For each $(i,j)$ node pair, we find all the shortest path nodes $W_{\mathcal{P}^*_{ij}}$. Nodes in $W_{\mathcal{P}^*_{ij}}$ are regarded as ground-truth set. 
    
    \item Obtain the geodesic $\gamma(i,j)$ and compute the distance to the geodesic $d(q, \gamma(i,j))$ for all nodes, excluding nodes $i$ and $j$.
    
    \item Determine the set of $\mathcal{S}$ nodes with the smallest distance to geodesic values, such that $|\mathcal{S}| =|W_{\mathcal{P}^*_{ij}}|$.
    
    \item Compute the ${\rm Precision} = |\mathcal{S}\bigcap W_{\mathcal{P}^*_{ij}} |/ |\mathcal{S}|$.
    
\end{enumerate}

\section{Results}\label{Sec:Results}
\subsection{Quantifying the Alignments of Shortest Paths Along Geodesic Curves}
\label{sec:Quantifying the alignment of shortest paths along geodesic curves in SRGG models}
\begin{figure}[ht]
    \centering
    \includegraphics[width=1\linewidth]{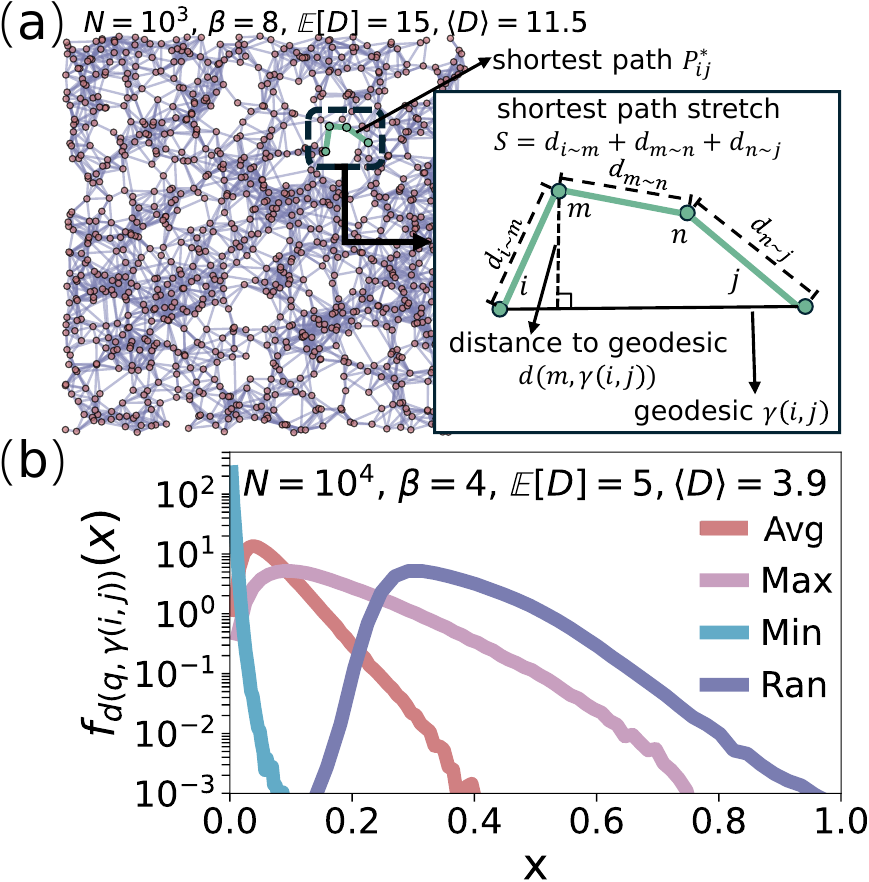}
    \caption{\footnotesize (a) A toy example of a geodesic, distance from a node to the geodesic and a shortest path (highlighted by green) with the corresponding shortest path stretch indicated by the black dashed line along the green path.
    (b) The distribution of the average (avg), maximum (max) and minimum (min) distances between shortest path nodes and geodesics connecting shortest path endpoints. We compare the resulting distributions to (ran) those between randomly chosen nodes and the geodesic. 
    Within an SRGG $G$, we randomly select $10^{6}$ shortest path endpoint pairs and for each $(i,j)$ pair, we find shortest path set $\mathcal{P}_{ij}^*$, geodesic $\gamma(i,j)$. 
    For shortest path nodes in $\mathcal{P}_{ij}^*$, we compute the minimum (min), maximum (max) and average (avg) distance to the geodesic. In addition, for each shortest path set $\mathcal{P}_{ij}^*$, we created a random set $\mathcal{P}_{ij}^{\rm rand}$ of the same cardinality containing randomly selected nodes. Using $\mathcal{P}_{ij}^{\rm rand}$, we computed average distances to the geodesic (ran).  All parameters used in the experiments are indicated in the respective panels.
    }
    \label{fig:deviation}
\end{figure}

We begin our work with the analysis of the alignment of shortest paths along geodesic curves in Soft Random Geometric Graphs (SRGGs). We generate an instance of a soft random geometric graph (SRGG) of $N=10^{4}$ nodes, connected with Fermi-Dirac connection probabilities given by Eq.~(\ref{eq:decreasing function for SRGG}) with inverse temperature $\beta = 4$, and parameter $d_0$ corresponding to the expected degree of $\mathbb{E}[D]=5$. Fig.~\ref{fig:deviation}(b) depicts the distributions of the maximum, the minimum, and the average distance to geodesic from shortest path nodes for $10^{6}$ randomly selected path endpoints within the SRGG, indicating that all three measures are substantially smaller than expected by chance. Our observation is not specific to the choice of SRGG parameters. We observe similar shortest path alignment properties in SRGGs with different parameters, Fig.~\ref{fig:DeviationDistributionN50_100}, and the same pattern persists when different link connection probability functions are used, Fig.~\ref{fig:waxman_distribution}.

After establishing that shortest path nodes are closer to geodesics than expected by chance, we next investigate how topological properties of  SRGGs affect path alignments. We measure the average distance to geodesic $\langle d \rangle$  and the average shortest path stretch $\langle S \rangle$, as described in Section~{\ref{sec:methods:alignment}}.
For each set of SRGG parameters, we generate $100$ SRGG realizations if $N\leq100$ and $1$ realization for $N > 100$.
For each generated graph $G_{i}$, $1 \leq i \leq 100$, we consider $1,000$ shortest paths between randomly selected connected node pairs when $N > 100$, and all connected node pairs in graphs with $N\leq100$.

\begin{figure*}[!]
    \centering
    \includegraphics[width=0.9\textwidth]{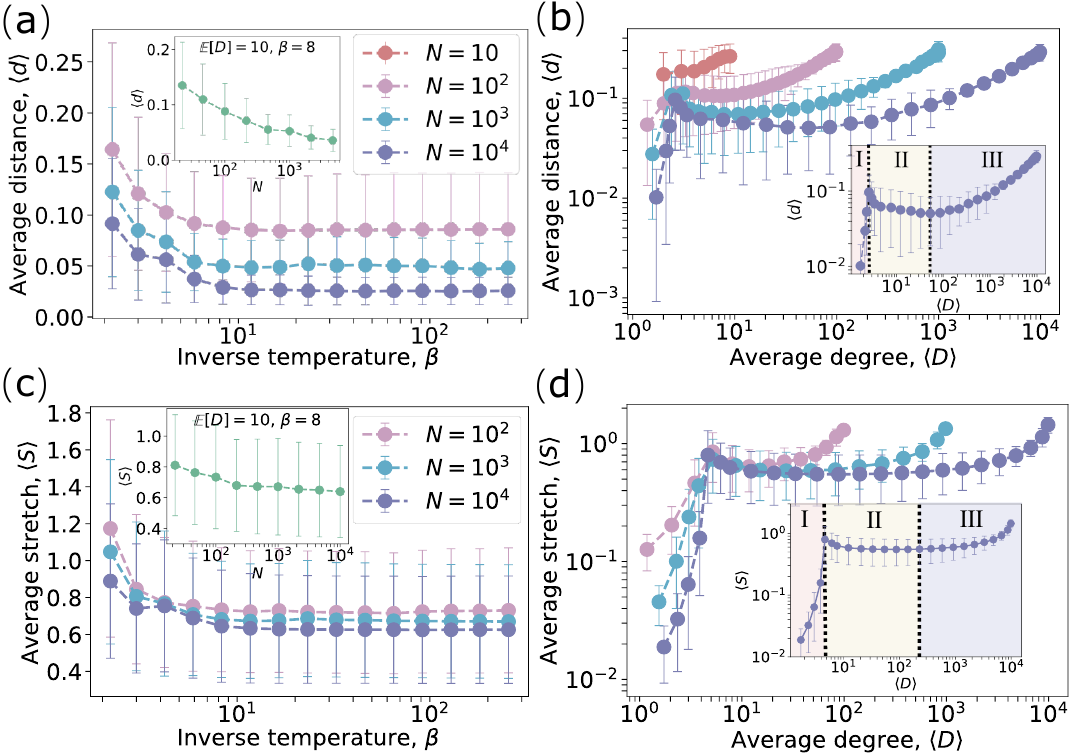} 
    \caption{\footnotesize (a) The average distance to the geodesic $\langle d \rangle$ as a function of inverse temperature $\beta$ for SRGGs with $\mathbb{E}[D] = 10$ and variable size $N$. The inset displays the average distance to the geodesic $\langle d \rangle$ as a function of the number of nodes $N$ for SRGGs with expected degree $\mathbb{E}[D] = 10$ and the inverse temperature $\beta=8$. (b) The average distance to the geodesic $\langle d \rangle$ as a function of the average degree $\langle D \rangle$ for SRGGs of variable size $N$ and the inverse temperature $\beta = 4$. The inset depicts the three phases discussed in the text. 
    (c) The average shortest path stretch $\langle S \rangle$ as a function of the inverse temperature $\beta$ for SRGGs of $\mathbb{E}[D] = 10$ and variable size $N$. The inset displays the average shortest path stretch $\langle S \rangle$ as a function of the number of nodes $N$ for SRGGs with expected degree $\mathbb{E}[D] = 10$ and the inverse temperature $\beta=8$. (d) The average shortest path stretch $\langle S \rangle$ as a function of the average degree $\langle D \rangle$ for SRGGs of variable size $N$ and the inverse temperature parameter $\beta = 128$. 
    All plots correspond to the average of $1,000$ randomly selected shortest paths, while the error bars quantify the standard deviations.
    }
    \label{fig:Dev_vs_diffpara}
\end{figure*}

Fig.~\ref{fig:Dev_vs_diffpara} depicts the average shortest path distance to geodesic $\langle d \rangle $ and the average shortest path stretch $\langle  S \rangle$ as a function of inverse temperature $\beta$ and the average degree $\langle D \rangle$. 

We observe that the average distance to the geodesic and the average path stretch decrease as $\beta$ increases, Fig.~\ref{fig:Dev_vs_diffpara}(a,c). This observation is expected. Indeed, by design, larger inverse temperature $\beta$ values favor short-distance connections in the SRGG, see Eq.~(\ref{eq:decreasing function for SRGG}). 
As a result, shortest path nodes are located closer to each other and closer to the geodesic connecting path endpoints, decreasing the average distance to the geodesic and the average shortest path stretch.

We also observe that the average distance to geodesic $\langle d \rangle $  and the average stretch $\langle S \rangle $  decrease as the network size $N$ increases, if the remaining SRGG parameters are held constant. 
Indeed, the density of nodes in the latent space increases as the network size $N$ increases.  Therefore, distances between connected nodes must decrease on average to preserve the expected degree $\mathbb{E}[D]$. It is the decrease of distances between connected nodes that leads to the decrease of the average distance to geodesic $\langle d \rangle $ and the average shortest path stretch $\langle S \rangle $, observed in Figs.~\ref{fig:Dev_vs_diffpara}(a,c).

The average distance to geodesics  $\langle d \rangle$ and the average path stretch $\langle S \rangle$  exhibit a non-monotonic behavior as a function of the graph average degree $\langle D \rangle$, Fig.~\ref{fig:Dev_vs_diffpara}(b,d). Initially, $\langle d \rangle$ and $\langle S \rangle$ increase as as the average degree $\langle D \rangle$ increases, Phase I in Fig.~\ref{fig:Dev_vs_diffpara}(b,d). After reaching the local maximum, $\langle d \rangle$ and $\langle S \rangle$ decrease as the average degree increases, Phase II, reaching minimum values $\langle d_{\rm min} \rangle$ $\langle S_{\rm min} \rangle$.  After reaching local minima, the average distance $\langle d \rangle$ and the average stretch  $\langle S \rangle$ resume their growth as a function of average degree in Phase III, see  Fig.~\ref{fig:Dev_vs_diffpara}(b,d).

\begin{figure*}[t]
    \centering
    \includegraphics[width=0.85\textwidth]{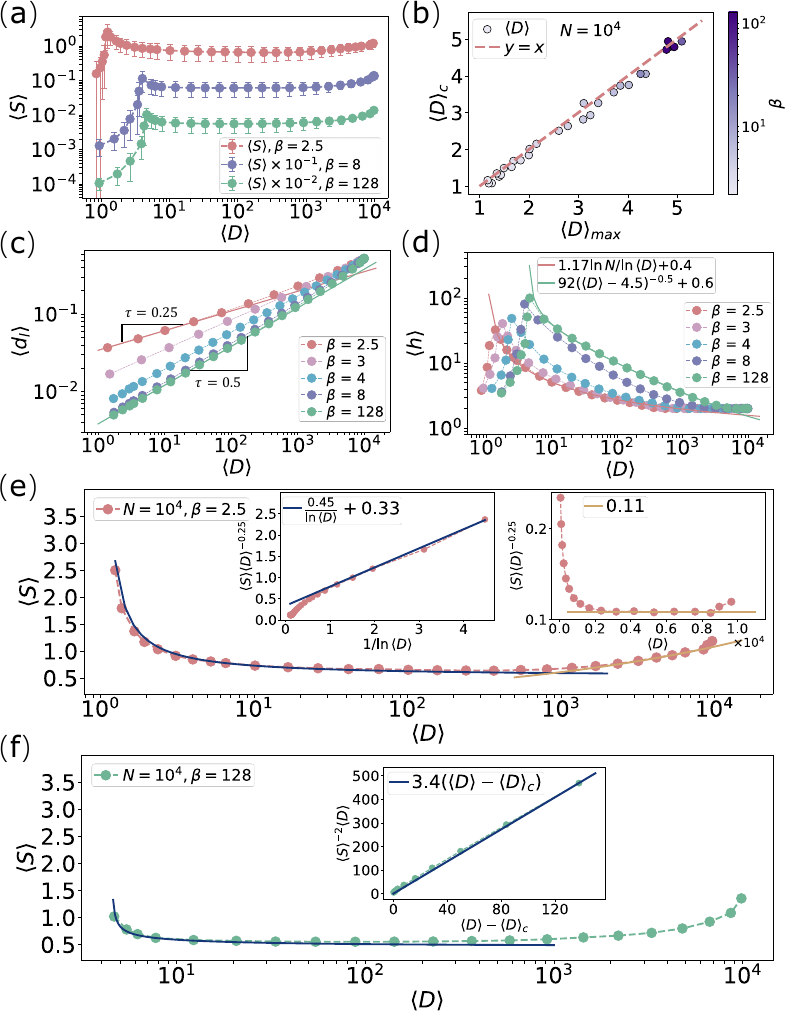}
    \caption{\footnotesize (a) The average shortest path stretch $\langle S \rangle$ as a function of the average degree $\langle D \rangle$ for inverse temperatures $\beta=1$, $\beta=8$, and $\beta=128$. The plots for beta $\beta = 8$ and $\beta=128$ are scaled by multiplicative factors $10^{-1}$ and $10^{-2}$, respectively, to avoid their overlap. The error bars depict standard deviation values.
    (b) Critical degree values $\langle D \rangle_c$ corresponding to the onset of the GCC as a function of expected degrees $\langle D\rangle_{\rm max}$ corresponding to the largest distance to geodesic. The scatter plot includes $20$ points corresponding to different inverse temperature values $\beta$ selected uniformly at random from the $(2,128]$ interval. 
    (c) The average link length $\langle d \rangle_{\ell}$ as a function of the average degree $\langle D \rangle$ in SRGGs with different inverse temperatures $\beta$. The solid lines are power-law curves with exponents $\tau = {\rm min} \left(1/2,\beta/2 - 1\right)$. 
    (d) The average shortest path hopcount $\langle h \rangle$ as a function of the average degree $\langle D \rangle$ in SRGGs with different inverse temperatures $\beta$. The solid lines are empirical fits for $\beta =2.5$ and $\beta=128$ values.
    (e,f) The average shortest path stretch $\langle S \rangle$ as a function of the average degree $\langle D \rangle$ for inverse temperatures (e) $\beta=2.5$ and (f) $\beta=128$. (e) The left inset depicts $\langle S \rangle \langle D \rangle^{-\tau}$ as a function of ${\rm ln} \langle D\rangle^{-1}$. Note that $\langle S \rangle \langle D \rangle^{-\tau} \propto {\rm ln} \langle D \rangle ^{-1}$, confirming the low-degree behavior of the shortest path stretch $\langle S \rangle$.  The right inset depicts  $\langle S \rangle \langle D \rangle^{-\tau}$  as a function of $\langle D \rangle$. Note that $\langle S \rangle \langle D \rangle^{-\tau}$ reaches a horizontal asymptote, confirming the power-law scaling of the shortest path stretch for large $\langle D \rangle$ values. (f) The inset depicts $\langle S \rangle^{-2} \langle D \rangle$ as a function of $\langle D \rangle -\langle D \rangle_c$, testing the low-degree behavior of the shortest path stretch. Note that the shortest path stretch $\langle S \rangle$ is nearly independent of $\langle D \rangle$ for a large range of $\langle D \rangle$ values.
}
    \label{fig:localminimum}
\end{figure*}

 We hypothesize that the initial growth of the average distance $\langle d \rangle$ and the average shortest path stretch $\langle S \rangle$ is related to the onset of the giant connected component (GCC) in the SRGGs. Indeed, when the average degree $\langle D \rangle$ is small, the network is fragmented into a large number of small disjoint subgraphs, Fig.~\ref{fig:Giant component with the expected degree}(b) in Appendix~\ref{app:Giant component with the expected degree}. 
As the average degree increases, disjoint subgraphs eventually merge into a giant connected component (GCC) of a size comparable to that of the entire graph. 
Right above the critical value of $\langle D \rangle = \langle D \rangle_c$ corresponding to the onset of the GCC, most of the nodes in the graph are connected, but the shortest paths between them may be arbitrarily long, see Fig.~\ref{fig:Giant component with the expected degree}(a) and Appendix~\ref{app:Giant component with the expected degree}.
As a result, the local maximum in the distance to geodesic $\langle d \rangle$ and the shortest path stretch $\langle S \rangle$ at the border of Phases I and II may coincide with the critical value $\langle D \rangle_c$ corresponding to the emergence of the GCC. 
To check if this is the case, we compared the $\langle D \rangle$  values corresponding to the local maximum of the distance to geodesic $\langle d \rangle$ with those corresponding to the GCC emergence for a rangle of SRGG parameters. As seen in Fig.~\ref{fig:localminimum}(b), the two values are nearly equal, leading to a strong correlation, Pearson $R=0.99$. The results support our hypothesis that the growth of the distance to geodesic $\langle d \rangle$ and the shortest path stretch $\langle S \rangle$ in Phase~I are driven by the formation of the giant connected component in SRGG.

After reaching the local maximum, the average path distance to geodesic $\langle d \rangle$ and the average path stretch $\langle S \rangle$ decrease as the average degree $\langle D \rangle$  increases until they reach their local minima at the border of Phase II and Phase III, Fig.~\ref{fig:Dev_vs_diffpara}(b,d). 

These non-monotonic behaviors of $\langle S \rangle$ and $\langle d \rangle$ are likely the results of the interplay of two competing effects. On the one hand, as the average degree increases, the average number of hops of a shortest path decreases, improving its alignment along a geodesic. On the other hand, larger average degrees result in larger distances between connected nodes, leading to larger deviations from the geodesic and larger stretch values.

To quantify these effects for the average shortest path stretch $\langle S \rangle$, we approximate it as a product of the average link length $\langle d_l \rangle$ and the average hopcount $\langle h \rangle$,
\begin{equation}
\langle S \rangle \approx \langle d_l \rangle \langle h \rangle.
\end{equation}

We estimate the average link length $\langle d_l \rangle$ using Bayes' rule. The conditional probability $\rho\left(D_{ij} = x |A_{ij} = 1\right)$ that the distance $D_{ij}$ between two randomly chosen nodes $i$ and $j$ equals $x$, given that the two nodes are connected, $A_{ij} = 1$, depends on the conditional probability nodes $i$ and $j$ are connected, given that the distance between them equals $x$, 
$P\left(A_{ij} = 1|D_{ij} = x\right)$, the probability density $\rho\left(D_{ij} = x\right)$ for a distance between two randomly chosen nodes in the unit square, and the probability $P\left(A_{ij} = 1\right)$ that two randomly chosen SRGG nodes are connected. Since $P\left(A_{ij} = 1|D_{ij} = x\right)$ is nothing else but the connection probability function $f(x)$ prescribed by Eq.~(\ref{eq:decreasing function for SRGG}), we have
\begin{equation}
\rho\left(D_{ij} = x |A_{ij} = 1\right) = \frac{f(x) \rho\left(D_{ij} = x\right)}{P\left(A_{ij} = 1\right)}.
\label{eq:dist_distances_srgg}
\end{equation}
Then the average link length $\langle d_l \rangle$ can be estimated, see Appendix~\ref{app:Derivation of the expected link length in SRGGs}, as 
\begin{equation}
\langle d_l \rangle =   \frac{\int x f\left(x\right) \rho\left(D_{ij} = x\right){\rm d}x}{\int f\left(x\right) \rho\left(D_{ij} = x\right){\rm d}x} \sim \langle D \rangle^{\tau},
\label{eq:exp_dist_srgg1}
\end{equation}
where $\tau = 1/2$ for $\beta > 3$ and  $\tau = \beta/2 -1$ for $\beta \in (2,3)$, see Fig.~\ref{fig:localminimum}(c). 

Fig.~\ref{fig:localminimum}(d) depicts the behavior of the average hopcount $\langle h\rangle$ as a function of the average degree $\langle D \rangle$. For large inverse temperatures $\beta$, SRGG is similar to an RGG, and we can approximate the behavior of the average hopcount as

\begin{equation}
\langle h \rangle \approx  \frac{\langle \delta \rangle}{\left[\langle D \rangle - \langle D \rangle_c\right]^{1/2}} + C,~~~\beta \gg 3
\label{eq:hopcount_large_beta}
\end{equation}
where $\langle \delta \rangle$ is the average distance between two randomly chosen points in the unit square,
$\langle D \rangle_c \approx 4.51$ is the critical value of the average degree corresponding to the onset of the GCC~\cite{penrose2003random}, and $C$ is a constant. For small inverse temperatures $\beta$, see Fig.~\ref{fig:localminimum}(d), SRGG is akin to a configurational model, and 
\begin{equation}
\langle h \rangle \approx \frac{{\rm ln} N}{{\rm ln} \langle D \rangle} + A,~~~\beta \in (2,3),
\label{eq:hopcount_small_beta}
\end{equation}
where $N$ is the number of nodes and for networks with Binomial degree distribution $A = \frac{1}{2}
- \frac{\gamma - 2 \log(1 - 1/\langle D \rangle)}{\log \langle D \rangle}
- \frac{2\,\mathbb{E}[\log W \mid W > 0]}{\log \langle D \rangle}.$ Here $\gamma$ is the Euler constant and $W$ is the almost sure limit of the normalized branching process~\cite{van2014performance,hooghiemstra2005mean}.
For graphs with expected degree $\mathbb{E}[D]>1.3$ and a binomial degree distribution, $\mathbb{E}[\log W \mid W > 0] \in (-0.2,0.2)$~\cite{van2014performance}. 
Since $\mathbb{E}[\log W \mid W > 0]$ is difficult to measure numerically, in this works we estimate the entire constant $A$ by fitting $\langle h \rangle$ vs $\langle D \rangle$ curves, see Fig.~\ref{fig:localminimum}(d).

The scaling behaviors of $\langle d_l \rangle$ and $\langle h \rangle$ given in Eqs.~(\ref{eq:exp_dist_srgg1}),~(\ref{eq:hopcount_large_beta}),~and~(\ref{eq:hopcount_small_beta})  account for the observed dependence of the average stretch $\langle S \rangle$ on the average degree $\langle D \rangle$ in phases II and III. 

For small inverse temperatures, $2<\beta<3$, the average stretch scales as
 $\langle S \rangle \propto \left(\frac{{\rm ln} N}{{\rm ln} \langle D\rangle} + C\right) \langle D\rangle ^{\frac{\beta}{2} - 1}$. Consequently, at small $\langle D\rangle$ (the onset of phase II), the stretch exhibits the asymptotic behavior $\langle S \rangle \sim \langle D\rangle^{\beta/2-1}/\ln \langle D\rangle$ [Fig.~\ref{fig:localminimum}(e)], whereas at large $\langle D\rangle$ (deep in phase III) the logarithmic correction becomes negligible and $\langle S \rangle \sim \langle D\rangle^{\beta/2-1}$, Fig.~\ref{fig:localminimum}.

For large inverse temperatures, $\beta \gg 3$, the average stretch follows $\langle S \rangle \approx \left(\frac{\langle \delta \rangle}{\left[\langle D \rangle - \langle D \rangle_c\right]^{1/2}} + C\right) \langle D\rangle ^{1/2}$.
As a consequence, for $\langle D\rangle \gg \langle D\rangle_c$ the dependence on $\langle D\rangle$ is weak, yielding stretch values that are approximately constant, as seen in Fig.~\ref{fig:localminimum}(f).

Overall, our results indicate that shortest paths in Soft Random Geometric Graphs are aligned along geodesic curves connecting shortest path endpoints. The alignment, as quantified by the average distance to geodesic $\langle d \rangle$ and the shortest path stretch $\langle S \rangle$, increases as the inverse temperature $\beta$ and the SRGG size $N$ increase. In contrast, the dependence of shortest-path alignment on the average degree $\langle D\rangle$ is nonmonotonic, reflecting the competition between two opposing effects: the reduction of the average hop count $\langle h\rangle$ and the concurrent increase of the average link length $\langle d_\ell\rangle$ as connectivity increases.

\subsection{Finding Shortest Path Nodes in Soft Random Geometric Graphs}
\label{sec:Identifying shortest path nodes in SRGG with noise}
\begin{figure*}[!]
    \centering
     \includegraphics[width = 1\textwidth]{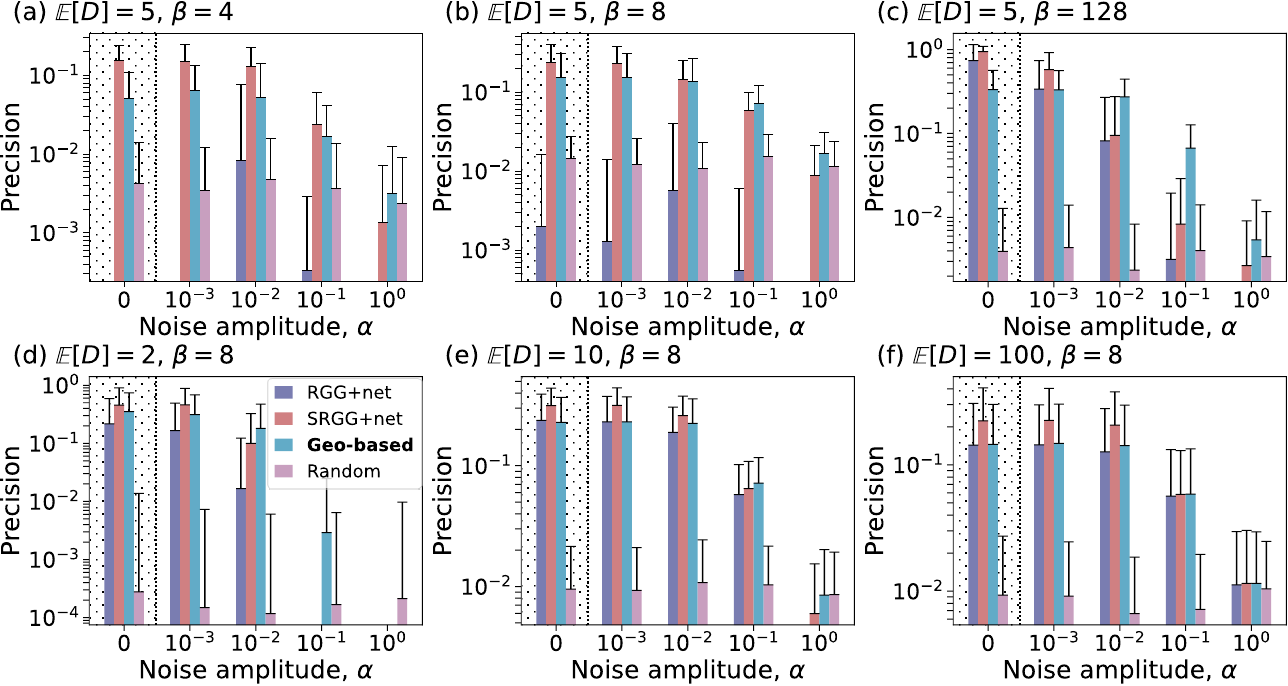}
    \caption{\footnotesize {\bf Precision of shortest path finding} using (i) distance to geodesic (Geo-based) and (ii) SRGG (SRGG+network) and (iii) RGG (RGG+network) reconstructions. Panels display average precision as a function of the coordinate noise amplitude $\alpha$ and panels correspond to different combinations of $\mathbb{E}[D]$ and $\beta$ parameters. All experiments are performed using SRGG networks of $N=10^{4}$ nodes. Each bar is the average precision computed for $100$ randomly chosen node pairs and the error bars depict standard deviations.
 }
    \label{fig:precision}
\end{figure*}

After observing and quantifying the alignment of shortest paths along geodesic curves in soft random geometric graphs, we next ask if shortest path nodes are identifiable based on their proximity to geodesic curves. To identify shortest path nodes connecting an $i-j$ node pair of interest, we draw a geodesic line $\gamma(i,j)$ connecting path endpoints $i$ and $j$ and then for every other network node $\ell \neq i\neq j$ we compute its distance to geodesic $d\left(\ell, \gamma(i,j)\right)$. The smaller the distance, the higher the likelihood for node $\ell$ to be a shortest path node, see Methods.

Fig.~\ref{fig:precision} displays the results of our path finding experiments for a combination of inverse temperature $\beta$ and expected degree $\mathbb{E}[D]$ parameters. We observe that Precision score for the distance to the geodesic in finding shortest path nodes is significantly larger than expected by chance.

Inspired by this observation, we ask if there are plausible circumstances in which the distance to the geodesic can find shortest-path nodes with accuracy higher than known state-of-the-art methods.  Clearly, if network $G$ is fully observable, existing graph theoretic algorithms, e.g., the Dijkstra algorithm, can find shortest path nodes precisely and in polynomial time. A much more challenging situation is when only network nodes and their approximate positions are known in network $G$. In more precise terms, we assumed that SRGG links are not known. At the same time, all SRGG nodes are known, but their coordinates are blurred with random noise of amplitude $\alpha > 0$:
\begin{equation}
\tilde{\mathbf{x}}_{i} = {\mathbf{x}}_{i} + \alpha X_i;\\
~\tilde{\mathbf{y}}_{i} =  {\mathbf{y}}_{i} + \alpha Y_i,
\label{eq:addnoise}
\end{equation}
where $i=1,\ldots,N$,  $X_i$ and $Y_i$ are random variables drawn from the uniform distribution $U[-1,1]$.

In this situation, one can still find shortest path nodes using the distance to the geodesic method with blurred node coordinates 
$\{\tilde{\mathbf{x}}_{i}, \tilde{\mathbf{y}}_{i}\}$. To find shortest path nodes with the Dijkstra network-based algorithm, one needs to first reconstruct SRGG links. To this end, we consider two strategies. The ${\rm SRGG}+{\rm net}$ strategy presumes that we know the connection probability function $f(x)$ used to generate the original SRGG graph. In this case, one first generates an SRGG instance with the same connection probability function $f(x)$ and then applies the Dijkstra algorithm to the resulting graph to find shortest path nodes. Another strategy, which we call ${\rm RGG}+{\rm net}$, assumes that the average degree $\langle D \rangle$ of the graph is known, but the connection probability function $f(x)$ is not. In this case, we treat the graph at hand as a random geometric graph, we compute its effective connectivity radius $R$ based on the number of nodes and the average degree, and then connect all nodes if their blurred positions are closer than $R$, see Methods (Section~\ref{sec:method predict}).

By comparing the precision of the distance to geodesics with those of the alternative ${\rm SRGG} + {\rm net}$ and ${\rm RGG} + {\rm net}$ strategies, we make several observations.  When node coordinates are known precisely, $\alpha = 0$, the ${\rm SRGG} + {\rm net}$ and ${\rm RGG} + {\rm net}$ strategies offer substantially higher Precision score than those of the distance to the geodesic, as shown in Fig.~\ref{fig:precision}. This result is expected since precisely known node coordinates allow for the reconstruction of graphs with relatively high accuracy. 

Second, both ${\rm SRGG} + {\rm net}$ and ${\rm RGG} + {\rm net}$ strategies become more accurate in finding shortest path nodes as the inverse temperature $\beta$ increases, Fig.~\ref{fig:precision}(a-c). We explain this observation by the decrease in randomness in SRGG networks: at higher $\beta$ values, connections are preferentially established at shorter distances, and distinct realizations of reconstructed networks are less different.  As a result, reconstructed networks $G_{\ell}$ better recreate the original network $G$, allowing for the identification of original shortest path nodes with higher precision, as shown in Fig.~\ref{fig:precision}.

Third, the ${\rm RGG} + {\rm net}$ strategy tends to be less accurate than the 
${\rm SRGG} + {\rm net}$. However, the relative accuracy of the former seems to increase as the inverse temperature $\beta$ increases. 
This result is also expected since larger $\beta$ values favor shorter links, effectively making SRGGs closer to RGGs. 

\begin{figure*}[!]
    \centering
     \includegraphics[width=1\textwidth]{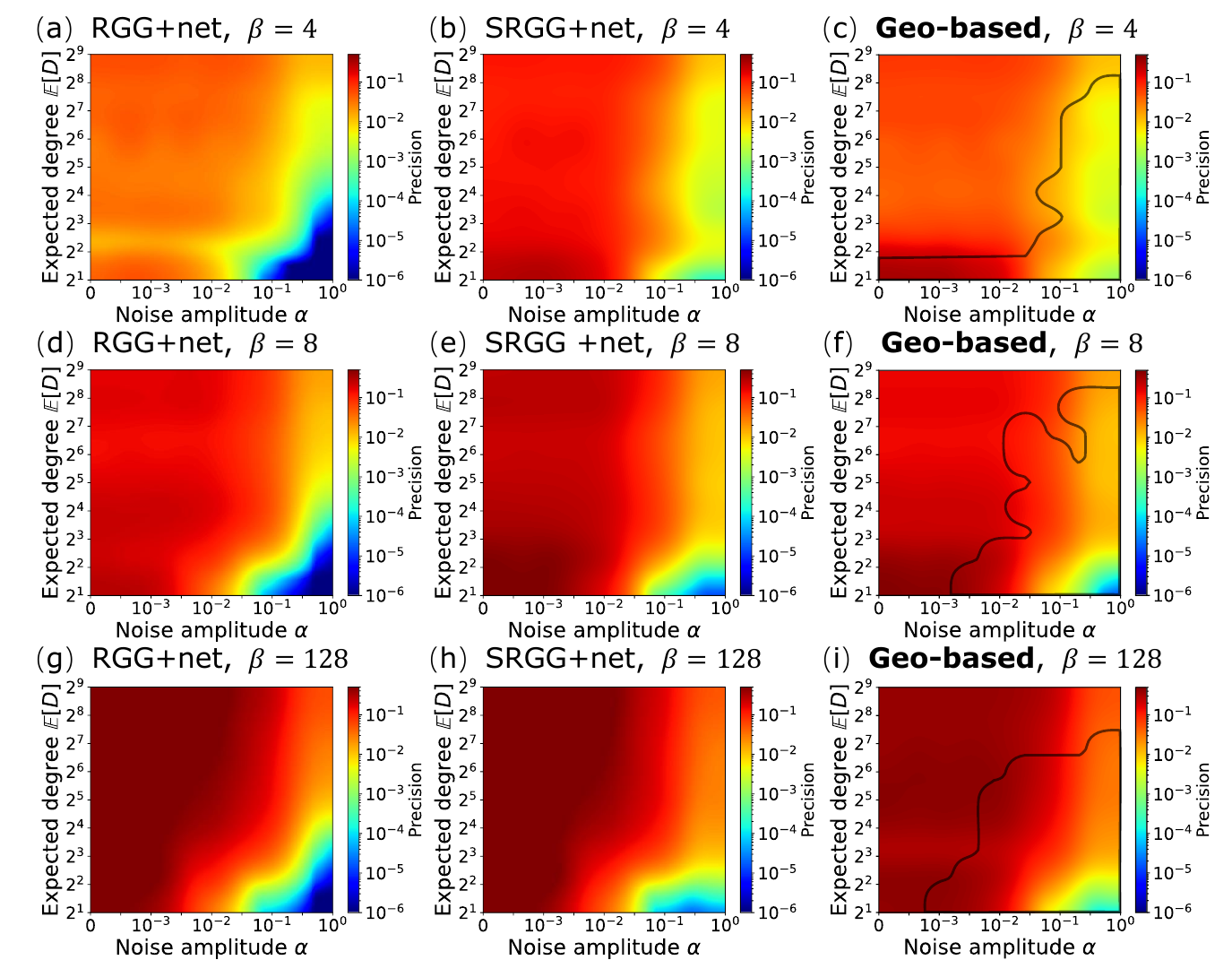}
    \caption{\footnotesize Heatmap of precision. The x-axis is the noise amplitude, while the y-axis represents the expected degree. 
    For each expected degree $\mathbb{E}[D]$ and inverse temperature $\beta$, a $10^4$-node SRGG is generated and the coordinates are respectively blurred with noise amplitude $\alpha$.
    The heatmaps consist of $9\times 9$ points, each point corresponding to the average of 100 randomly selected node pairs in a (blurred) SRGG. In the region enclosed by the black line, the distance to geodesic approach (Geo-based) achieves higher precision compared with the SRGG (SRGG+network) and RGG (RGG+network) reconstruction approaches.
 }
    \label{fig:PrecsionHeatmapwithnoise}
\end{figure*}

When comparing the distance to the geodesic approach to the ${\rm SRGG} + {\rm net}$ and ${\rm RGG} + {\rm net}$ strategies, we find that the precision of the distance to geodesic approach may exceed that of the ${\rm SRGG} + {\rm net}$ and ${\rm RGG} + {\rm net}$ when the noise amplitude $\alpha$ is large, Fig.~\ref{fig:PrecsionHeatmapwithnoise}, indicating that the distance to the geodesic approach is more robust to coordinate uncertainities than network reconstruction approaches ${\rm SRGG} + {\rm net}$ and ${\rm RGG} + {\rm net}$. In the case of noise magnitude $\alpha$, distances between nodes may be perturbed by at most $2\alpha$ and lead to changes in some node rankings with respect to distances to geodesics. 
The effects of node coordinates on network reconstruction approaches are potentially more disrupting. This is the case since link probabilities in the SRGG depend nonlinearly on distances between the nodes and even small changes in network links may significantly alter shortest paths. As a result, shortest paths in reconstructed networks may significantly differ from those in the original network, reducing the accuracy of network reconstruction strategies.

To explore the limits of shortest path finding with the distance to geodesic, we next conduct systematic experiments for ranges of noise amplitude $\alpha \in [0,1]$, average degree $\mathbb{E}[D] \in \left[2,2^{9}\right]$, and different inverse temperature values. As seen in Fig.~\ref{fig:PrecsionHeatmapwithnoise}, the accuracy of all path finding methods decreases as the average degree $\mathbb{E}[D]$ decreases and coordinate uncertainty $\alpha$ increases. At the same time, the accuracy of the distance to the geodesic approach is less sensitive to noise, resulting in low average degree and high noise regions, Fig.~\ref{fig:PrecsionHeatmapwithnoise}(c), (f), and (g), where the distance to the geodesic method is more accurate than network reconstruction strategies.

\section{Discussion}
\label{sec:Conclusion}

In summary, we analyzed the alignment of shortest paths along geodesic lines connecting shortest path endpoints in Soft Random Geometric Graphs (SRGGs). Some of our results are intuitive and expected. In particular, we found that the alignments of shortest paths, as measured by the distance to the geodesic, and the path stretch become stronger in SRGGs as the number of nodes $N$ and the inverse temperature $\beta$ increase.

Nevertheless, the non-monotonic behavior of shortest path alignment as a function of the average degree is far from trivial. As the average degree increases, two competing phenomena take place. On the one hand, the hopcount of shortest paths decreases, limiting the extent of {\it wandering} of shortest paths. On the other hand, larger average degree values can only be achieved by allowing nodes to connect over larger {\it Euclidean} distances. As a result, distances between adjacent nodes in a shortest path increase, allowing for large distances from shortest path nodes to geodesic curves. The competition between these two effects results in the non-monotonic behavior of the distance to the geodesic as a function of the average degree.

We found that the alignments of shortest paths along geodesic curves in {\it Euclidean} SRGGs are sufficiently strong to enable shortest path finding, even near criticality where deviation measures such as $\langle S\rangle$ and $\langle d\rangle$ reach their maximum values. Shortest path nodes in {\it Euclidean} SRGGs can be identified using distance to geodesic, similar to random hyperbolic graphs~\cite{kitsak2023finding}. 
The accuracy of such geometric shortest path finding is higher for graphs with a larger inverse temperature $\beta$, in agreement with our path alignment experiments.

The distance to the geodesic method uses node coordinates to find nodes closest to the geodesic connecting shortest path endpoints, and can be used in situations when node links are not known. In contrast, to enable traditional network-based methods in this situation, one needs first to reconstruct network links, which  
can be a highly non-trivial task. We find that the best advantage of the distance to the geodesic method is achieved for small average degree values and larger node coordinate uncertainties. While the geometric alignment of shortest path nodes is not necessarily the strongest in this case, network reconstruction methods are even less accurate, offering better prospects for using network geometry in finding shortest path nodes.

Our results are not specific to the choice of the connection probability function. While most of our analysis is based on SRGGs with the Fermi-Dirac connection probability function, Eq.~(\ref{eq:decreasing function for SRGG}), we obtain qualitatively similar results for the alignment of shortest paths for Waxman SRGGs characterized by the exponentially decaying connection probabilities, Eq.~(\ref{eq:waxman_conn}), see Fig.~\ref{fig:waxman_distribution}. 

\begin{figure}
    \centering
    \includegraphics[width=1\linewidth]{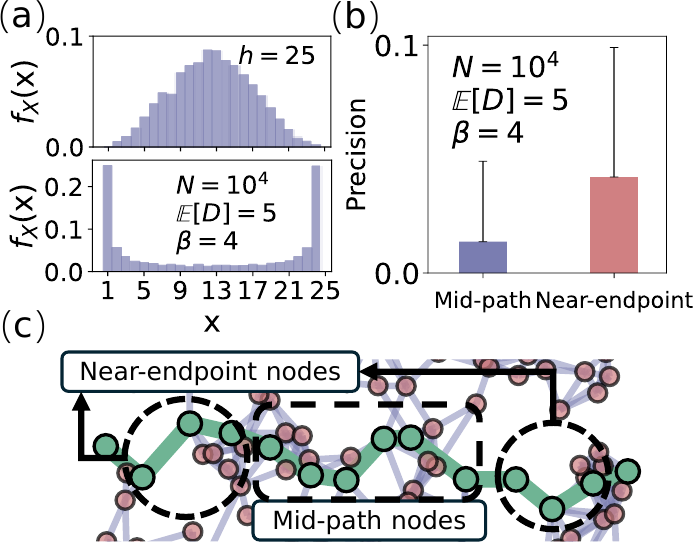}
    \caption{ \footnotesize (a, top) The distribution of shortest path node positions  with the largest distance to geodesic. (a, bottom) The distribution of shortest path node positions with the smallest distance to geodesic. In both distributions, we consider shortest paths of fixed hopcount, $h = 25$, and SRGGs with $N=10^{4}$ nodes, inverse temperature $\beta = 4$, and the expected degree $\mathbb{E}[D]=5$. The most remote from the geodesic nodes are likely to be found in the middle of a shortest path, while the closest to the geodesic nodes are likely to be found close to the path endpoints. See Appendix~\ref{app:Deviation} for plots with other SRGG parameters.
    (b) The precision of finding shortest path nodes near endpoints (Near-endpoint) and in the middle of the path (Mid-path) using the distance to the geodesic.
    A $10^4$-node SRGG is generated.
    Each bar is the average precision computed for $100$ randomly selected node pairs and the error bars depict standard deviations.
    (c) shows an example illustrating the division of Near-endpoint nodes and Mid-path nodes on a shortest path.}
    \label{fig:predictcentersidenode}
\end{figure}

Our findings invite follow-up questions. One of them, inspired by real networks, is the problem of shortest path finding in geometric networks where nodes are distributed heterogeneously. While the distance to the geodesic can also be used in geometric networks with heterogeneous spatial node distributions, it is not clear if the accuracy of path-finding remains high. 

Another open question outside the scope is the geometric properties of shortest paths in weighted geometric networks. While link weights can be functions of links distances, more elaborate models were proposed in the literature that decouple links weights from distances between connected nodes~\cite{simini2012universal,allard2017geometric,voitalov2020weighted}. 

Finally, the important question is whether all shortest path nodes are equally difficult to find using distance to the geodesic. Our preliminary experiments indicate that shortest path nodes situated closer to path endpoints tend to have a smaller distance to the geodesic and, as a result, are easier to infer, Fig.~\ref{fig:predictcentersidenode}. This observation is exciting since shortest path nodes closer to the path endpoints may be easier to manipulate in networks compared to nodes in the middle of the shortest path. Indeed, nodes in the middle of a shortest path may be shared by other paths and can be involved in many vital processes. As a result, altering their connections or capacity may lead to cascades of changes throughout the network~\cite{PhysRevE.97.012309, yang2017small, lu_vital_2016, qiu2021identifying}.

Figure~\ref{fig:predictcentersidenode} suggests that shortest path nodes closer to path endpoints are easier to find with the distance to the geodesic. {\it Are nodes in the middle of a shortest path easier to find with network-based methods?} This can indeed be the case since (i) nodes in the middle of a shortest path tend to have degrees larger than those close to its endpoints and (ii) large degree nodes are often characterized by large betweenness centrality values~\cite{freeman1977set, PhysRevE.75.056115}. As a result, nodes in the middle of shortest paths can be identified by prioritizing large-degree nodes. Thus, a fusion approach combining elements of the distance to geodesic with network topology may ultimately be more accurate in finding shortest path nodes than one of the two approaches used separately.

\begin{acknowledgments}
M. Kitsak has been supported by the Dutch Research Council (NWO) grants OCENW.M20.244 and VI.C.242.106. P. Van Mieghem is supported by the European Research Council under the European Union’s Horizon 2020 research and innovation program (Grant Agreement 101019718).
Z. Qiu has been funded by the China Scholarship Council (Grant No. 202106070032)
\end{acknowledgments}

\begin{widetext}
\appendix
\renewcommand{\thefigure}{\thesection\arabic{figure}}
\newpage
\section{Notation}
\setlength{\arrayrulewidth}{0.6pt}
\begin{table}[h]
	\centering
	\caption{Notation}
	\label{symbol}
	\begin{tabular}{ll}
		\hline
		Symbol& Definition\\
		\hline
		$G$& Graph or network\\
		$N$& Number of nodes\\
		$L$& Number of links \\
            $l= i\sim j$& A link between nodes $i$ and $j$\\
            $C_G$& The average clustering coefficient of graph $G$ \\
            $\mathbb{E}[D]$ & The expected degree of a graph\\
            $\langle D \rangle$ & The average degree of a graph\\
            $\mathcal{P}_{ij}$& A path from node $i$ to node $j$\\
		$\mathcal{P}^*_{ij}$& A shortest path between nodes $i$ and $j$\\
        $W_{\mathcal{P}^*_{ij}}$ & Nodes contained in all shortest paths connecting nodes $i$ and $j$\\
		$h$& The hopcount \\
		$\gamma(i,j)$ & The geodesic curve between nodes $i$ and $j$\\
            $d_{ij}$ & The geometric distance between nodes $i$ and $j$\\
            $d_l$ & The geometric length of link $l$, i.e., the distance between its two adjacent nodes\\
            $d(q,\gamma(i,j))$& The geometric distance from node $q$ to the geodesic $\gamma(i,j)$\\
            $\langle d\rangle$ & The average distance to the geodesic\\
            $S$ & The shortest path stretch, i.e., the sum of its link lengths. \\
		\hline
	\end{tabular}
\end{table}

\section{Distance to geodesic}
\setcounter{figure}{0}
\label{app:Deviation}
\begin{figure*}[!b]
    \centering
     \includegraphics[width = 0.8\linewidth]{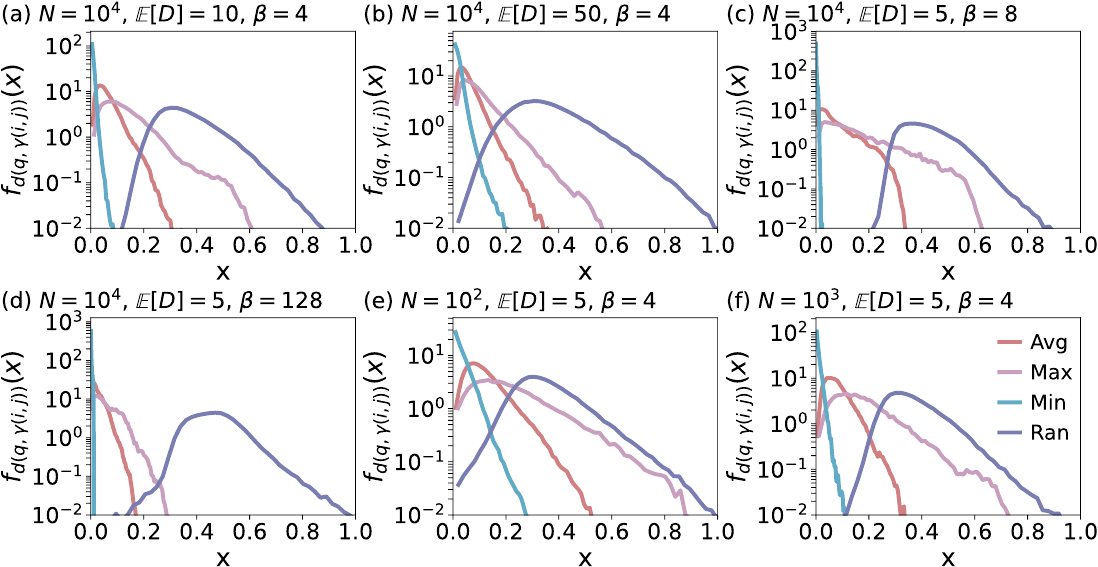}
    \caption{\footnotesize The distribution of the average (avg), maximum (max), and minimum (min) distances between shortest path nodes and geodesics connecting shortest path endpoints. For comparison, we also plot the distribution of distances to geodesic from randomly selected nodes (ran). The distributions in (a)-(d) are obtained for a single SRGG realization and $10^6$ random shortest paths.
    For panels (e) and (f), we generated $10^3$ SRGG realizations. For each $G_{i}$, $i = 1,\ldots, 10^{3}$, we considered $10^3$ random shortest paths. 
    }
    \label{fig:DeviationDistributionN50_100}
\end{figure*}
Figure~\ref{fig:DeviationDistributionN50_100} depicts the distributions of the distance to the geodesic in SRGGs for a variety of expected degree $\mathbb E[D]$ and inverse temperature $\beta$ values.
Figure~\ref{fig:MAXDeviationNodeDistribution} demonstrates the distribution of the node positions $X$ in the shortest path corresponding to the largest distance to the geodesic.
Figure~\ref{fig:MinDeviationNodeDistribution} demonstrates the distribution of the node positions $X$ in the shortest path corresponding to the minimum distance to the geodesic.

In this section, we also study the distributions of the distance to the geodesic for Waxman SRGGs with the connection probability function 
\begin{equation}
f(d) = \beta e^{\left(-\frac{d}{d_0}\right)^{\eta}}. 
\end{equation}
Waxman SRGGs of $N$ nodes built on the unit square with uniform node density are characterized by the expected degree
\begin{equation}
\mathbb{E}[D]\approx 2\pi N \beta \frac{d_0^{2}}{\eta}
\Gamma\!\left(\frac{2}{\eta}\right).
\end{equation}

Figure~\ref{fig:waxman_distribution} depicts the distributions of the distance to the geodesic for a variety of expected degree $\mathbb E[D]$ and parameter $\eta$ values. Note that the results for the Waxman SRGGs are similar to SRGGs with the Fermi-Dirac connection probability function.

\begin{figure*}[!h]
    \centering
    \includegraphics[width = 0.85\textwidth]{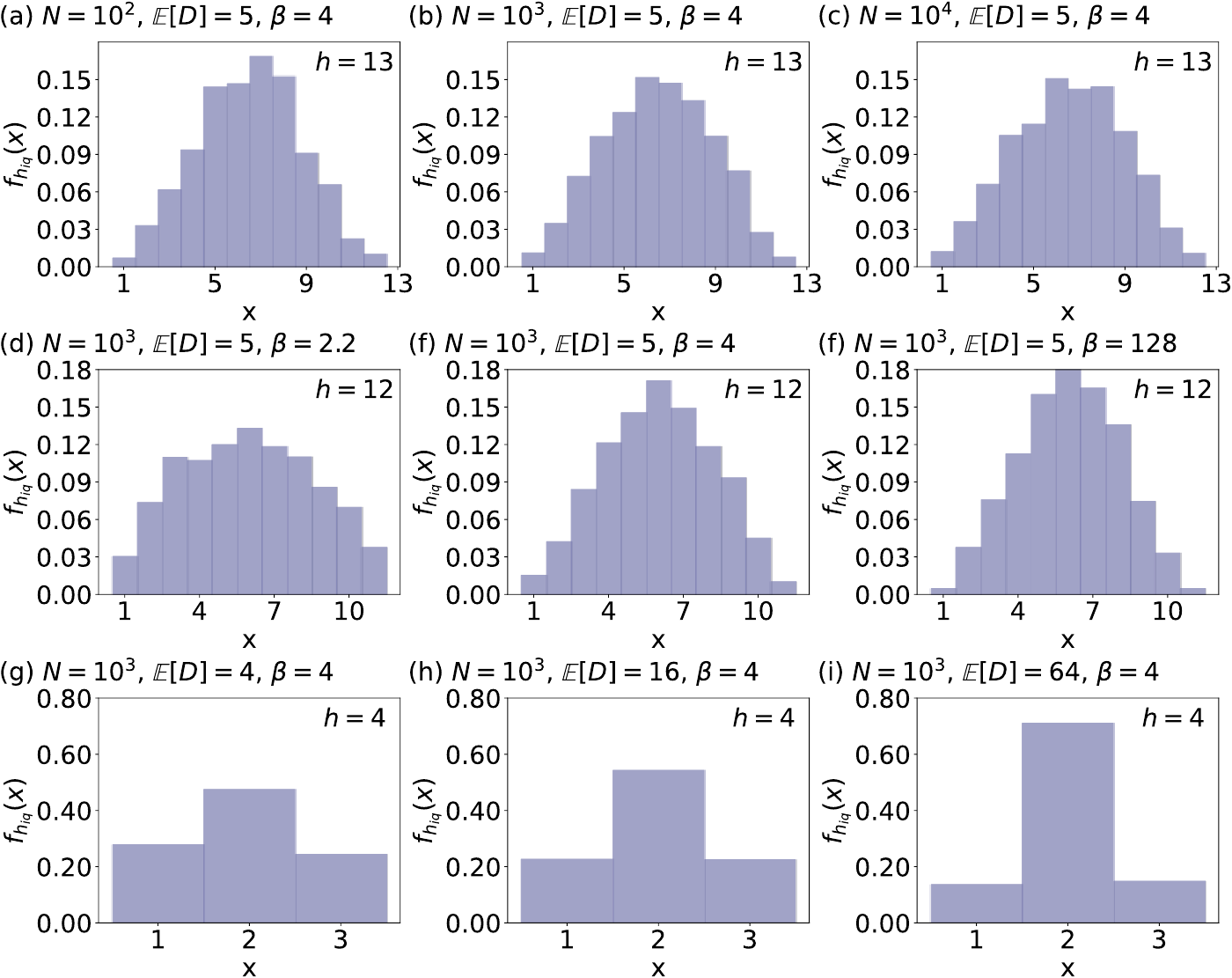}
    \caption{\footnotesize  
    Distribution of the position $X$ of the node with the largest distance to the geodesic along shortest paths in SRGGs for different combinations of network size $N$, inverse temperature $\beta$, and the expected degree $\mathbb{E}[D]$. In each panel, only shortest paths with a fixed hop count $h$ are considered. For each shortest path, the node with the maximum distance to the geodesic is identified, and its position $X \in \{1,\ldots,h\}$ along the path is recorded. Panel (a) averages over $10^{3}$ independent SRGG realizations, while the remaining panels correspond to a single SRGG realization. In all cases, statistics are collected from $10^{5}$ randomly selected shortest paths per SRGG realization.}
    \label{fig:MAXDeviationNodeDistribution}
\end{figure*}

\begin{figure*}[!h]
    \centering
    \includegraphics[width = 0.85\textwidth]{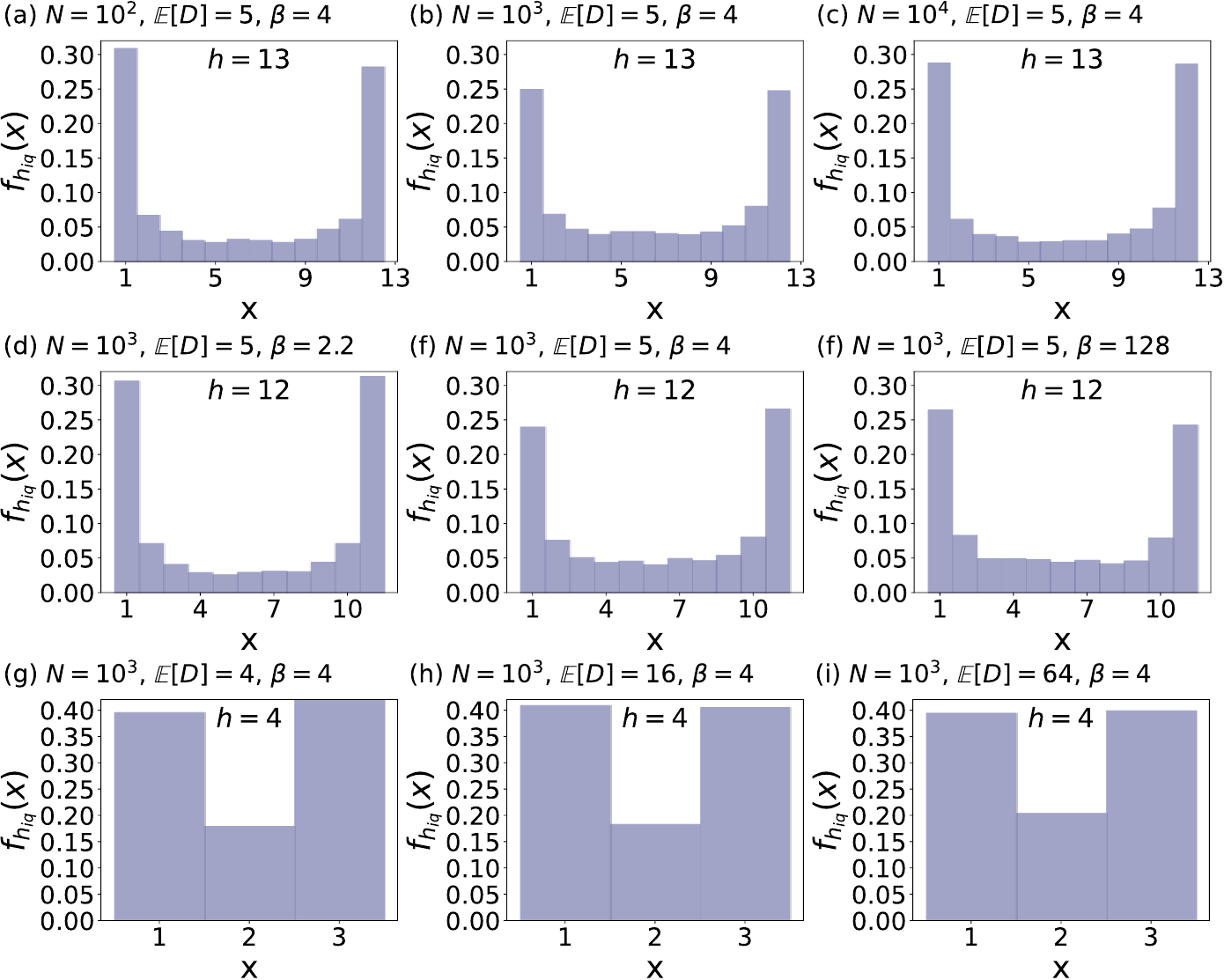}
    \caption{\footnotesize  
    Distribution of the position $X$ of the node with the smallest distance to the geodesic along shortest paths in SRGGs for different combinations of network size $N$, inverse temperature $\beta$, and the expected degree $\mathbb{E}[D]$. In each panel, only shortest paths with a fixed hop count $h$ are considered. For each shortest path, the node with the minimum distance to the geodesic is identified, and its position $X \in \{1,\ldots,h\}$ along the path is recorded. Panel (a) averages over $10^{3}$ independent SRGG realizations, while the remaining panels correspond to a single SRGG realization. In all cases, statistics are collected from $10^{5}$ randomly selected shortest paths per SRGG realization.}
    \label{fig:MinDeviationNodeDistribution}
\end{figure*}

\begin{figure*}[!h]
    \centering
    \includegraphics[width = 0.85\textwidth]{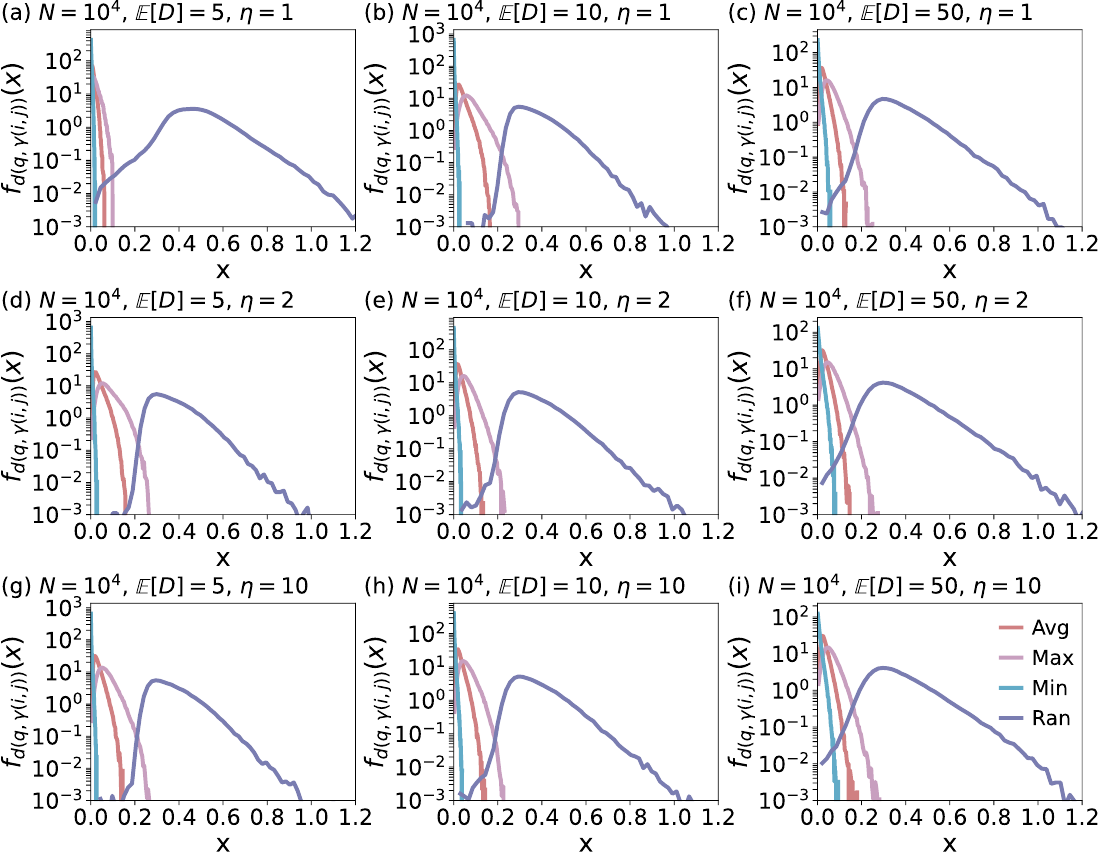}
    \caption{\footnotesize  
    The distribution of the average (avg), maximum (max), and minimum (min) distances between shortest path nodes and geodesics connecting shortest path endpoints in Waxman SRGGs with the connection probability function given by Eq.~\eqref{eq:waxman_conn}. For comparison, we also plot the distribution of distances to geodesic from randomly selected nodes (ran). The network size $N=10^4$ and the parameter $\beta = 1$.
    For each expected degree $\mathbb{E}[D]$ and parameter $\eta$, we generated an SRGG and considered $10^6$ random shortest paths.}
    \label{fig:waxman_distribution}
\end{figure*}

\section{The onset of the giant connected component in the SRGG}
In Fig.~\ref{fig:Giant component with the expected degree}, we show the onset of the giant connected component (GCC) in the SRGGs.
\label{app:Giant component with the expected degree}
\setcounter{figure}{0}
\begin{figure}[!h]
    \centering
    \includegraphics[width = 0.5\linewidth]{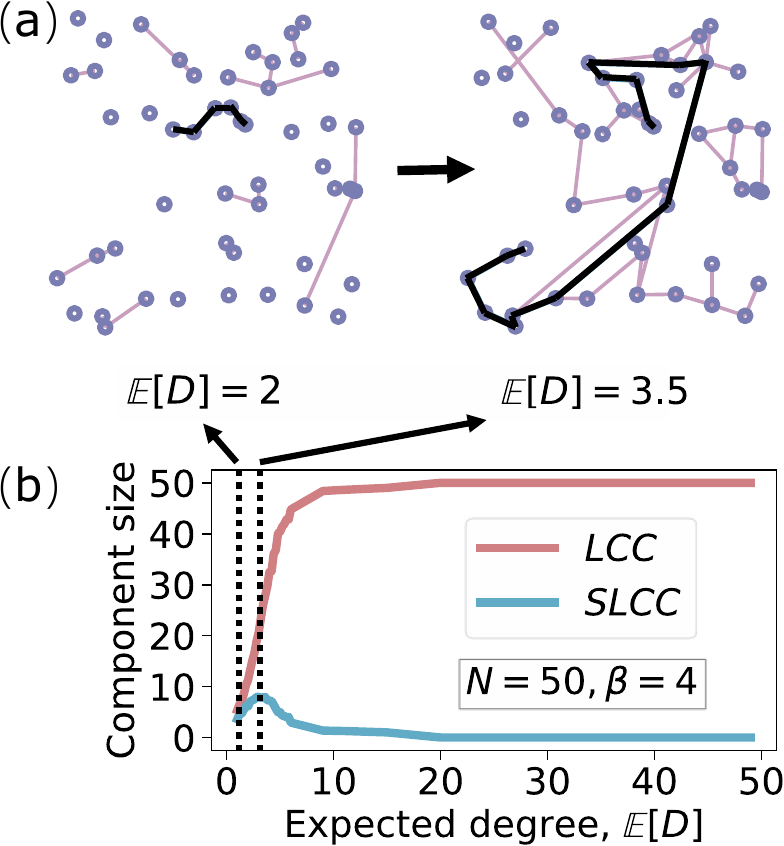}
    \caption{\footnotesize (a) Visualizations of toy SRGGs with $N=50$ nodes and inverse temperature $\beta=4$ for increasing expected degree, from left to right:
$\mathbb{E}[D]=2$, $\mathbb{E}[D]=3.5$.
In each network, a representative shortest path is highlighted in green.
(b) Average sizes of the largest connected component (LCC) and the second largest connected component (SLCC) for SRGGs with $N=50$ and $\beta=4$ as a function of the expected degree $\mathbb{E}[D]$.
The emergence of a giant connected component occurs at the critical expected degree $D_c=3.5$, identified by the maximum of the SLCC.
All LCC and SLCC values are averaged over 100 independent realizations.
Accordingly, the network visualizations in panel (a) correspond to the disconnected and critical connected regimes.}
    \label{fig:Giant component with the expected degree}
\end{figure}

\section{The average link length}
\label{app:Derivation of the expected link length in SRGGs}

To derive the expression for the average link length $\langle d_{l} \rangle$ in SRGG models, we first define $\rho\left(D_{ij} = x |A_{ij} = 1\right)$ as the conditional probability density that the distance $D_{ij}$ between two randomly chosen nodes in the SRGG equals $x$, provided these nodes are connected, $A_{ij} = 1$. The average link length is then given by $\langle d_{l} \rangle = \int_{0}^{\sqrt{2}} x \rho\left(D_{ij} = x |A_{ij} = 1\right){\rm d} x$.

Using the Bayes' rule, the thought  probability density function can be expressed as
\begin{equation}
\rho\left(D_{ij} = x |A_{ij} = 1\right) = \frac{P\left(A_{ij} = 1|D_{ij} = x\right) \rho\left(D_{ij} = x\right)}{P\left(A_{ij} = 1\right)},
\label{eq:dist_distances_srgg2}
\end{equation}
where $P\left(A_{ij} = 1|D_{ij} = x\right)$ is nothing else but the connection probability function $f(x)$, prescribed by Eq.~(\ref{eq:decreasing function for SRGG}), and $\rho\left(D_{ij} = x\right)$ is the distribution of distances between node pairs in a unit square. $P\left(A_{ij} = 1\right)$ is the probability that a randomly chosen node pair is connected, $P\left(A_{ij} = 1\right) = \frac{2 L}{N(N-1)} = \frac{\langle D \rangle} {N-1}$.

The average link length is then given by
\begin{align}
\langle d_l \rangle &= \frac{\int_{0}^{\sqrt{2}}  x f\left(x\right) \rho\left(D_{ij} = x\right){\rm d} x}{\int_{0}^{\sqrt{2}}  f\left(x\right) \rho\left(D_{ij} = x\right){\rm d} x}.
\label{eq:exp_dist_srgg2}
\end{align}
We approximate the distribution of distances between
node pairs in a unit square as $\rho\left(D_{ij} = x\right) = \frac{2x}{R^2}$, where $R$ is the size of a unit square, $R = \mathcal{O}(1)$, obtaining
\begin{align}
\langle d_l \rangle 
& \approx d_0\frac{\int_{0}^{(\sqrt{2}/d_0)^\beta}  \frac{u^{3/\beta-1}}{1+ u}{\rm d} u}{\int_{0}^{(\sqrt{2}/d_0)^\beta}  \frac{u^{2/\beta-1}}{1+ u}{\rm d} u}.
\label{eq:exp_dist_srgg3}
\end{align}

The closed-form expression for $\langle d_l \rangle$  in Eq.~(\ref{eq:exp_dist_srgg3}) is given by the ratio of two hypergeometric functions. It is useful, however, to infer its leading order behavior as a function of $\langle D \rangle$. To this end, we recall that $\frac{1}{d_0} \sim N^{1/2} \gg 1$ for large $N$. This follows from Eq.~(\ref{eq:exp_degree_srgg}) for sparse SRGGs. 

Since $\beta > 2$, the integral in the denominator of Eq.~(\ref{eq:exp_dist_srgg3}) can be approximated as ${\int_{0}^{\infty}  \frac{u^{2/\beta-1}}{1+ u}{\rm d} u} = \pi / {\rm sin} \left(\frac{2 \pi}{\beta}\right)$. Further, if $\beta > 3$, the integral in the numerator of  Eq.~(\ref{eq:exp_dist_srgg3}) can be approximated in a similar manner as ${\int_{0}^{\infty}  \frac{u^{3/\beta-1}}{1+ u}{\rm d} u} = \pi / {\rm sin} \left(\frac{3 \pi}{\beta}\right)$. As a result,
\begin{equation}
\langle d_l \rangle \approx d_0\frac{{\rm sin} \left(\frac{2 \pi}{\beta}\right)}{{\rm sin} \left(\frac{3 \pi}{\beta}\right)} \sim \langle D\rangle ^{1/2},~~~\beta>3.
\end{equation}

In case $\beta \in (2,3)$, we approximate the integral in the numerator of Eq.~(\ref{eq:exp_dist_srgg3}) as ${\int_{0}^{\left(\sqrt{2}/d_0\right)^{\beta}}  u^{3/\beta-2}{\rm d} u}$, which results in 

\begin{equation}
\langle d_l \rangle \approx \frac{ \pi \beta 2^{\frac{3-\beta}{2}}{\rm sin} \left(\frac{2 \pi}{\beta}\right)}{3-\beta} {d}^{\beta-2}_0 \sim \langle D\rangle ^{\beta/2-1},~~~2<\beta<3.
\end{equation}
\end{widetext}

\newpage
\bibliographystyle{apsrev4-2}
\bibliography{apssamp}

@PREAMBLE{
 "\providecommand{\noopsort}[1]{}" 
 # "\providecommand{\singleletter}[1]{#1}%" 
}

@article{waxman2002routing,
  title={Routing of multipoint connections},
  author={Waxman, Bernard M},
  journal={IEEE Journal on Selected Areas in Communications},
  volume={6},
  number={9},
  pages={1617--1622},
  year={2002},
  doi = {10.1109/49.12889},
  publisher={IEEE}
}

@article{yang2017small,
  title={Small vulnerable sets determine large network cascades in power grids},
  author={Yang, Yang and Nishikawa, Takashi and Motter, Adilson E},
  journal={Science},
  volume={358},
  number={6365},
  pages={eaan3184},
  year={2017},
    doi = {10.1126/science.aan3184},
  publisher={American Association for the Advancement of Science}
}

@article{PhysRevE.75.056115,
  title = {Betweenness centrality of fractal and nonfractal scale-free model networks and tests on real networks},
  author = {Kitsak, Maksim and Havlin, Shlomo and Paul, Gerald and Riccaboni, Massimo and Pammolli, Fabio and Stanley, H. Eugene},
  journal = {Physical Review E},
  volume = {75},
  issue = {5},
  pages = {056115},
  numpages = {8},
  year = {2007},
  month = {May},
  publisher = {American Physical Society},
  doi = {10.1103/PhysRevE.75.056115},
}

@article{qiu2021identifying,
  title={Identifying vital nodes by Achlioptas process},
  author={Qiu, Zhihao and Fan, Tianlong and Li, Ming and L{\"u}, Linyuan},
  journal={New Journal of Physics},
  volume={23},
  number={3},
  pages={033036},
  year={2021},
    doi = {10.1088/1367-2630/abe971},
  publisher={IOP Publishing}
}

@article{PhysRevE.97.012309,
  title = {Stability of a giant connected component in a complex network},
  author = {Kitsak, Maksim and Ganin, Alexander A. and Eisenberg, Daniel A. and Krapivsky, Pavel L. and Krioukov, Dmitri and Alderson, David L. and Linkov, Igor},
  journal = {Physical Review E},
  volume = {97},
  issue = {1},
  pages = {012309},
  numpages = {7},
  year = {2018},
  month = {Jan},
  publisher = {American Physical Society},
  doi = {10.1103/PhysRevE.97.012309},
}

@article{freeman1977set,
  title={A set of measures of centrality based on betweenness},
  author={Freeman, Linton C},
  journal={Sociometry},
  pages={35--41},
  year={1977},
doi = {10.2307/3033543},
  publisher={JSTOR}
}

@article{gomathi2018energy,
  title={Energy efficient shortest path routing protocol for underwater acoustic wireless sensor network},
  author={Gomathi, RM and Martin Leo Manickam, J},
  journal={Wireless Personal Communications},
  volume={98},
  pages={843--856},
  year={2018},
doi = {10.1007/s11277-017-4897-5},
  publisher={Springer}
}

@article{chen2012reliable,
  title={Reliable shortest path finding in stochastic networks with spatial correlated link travel times},
  author={Chen, Bi Yu and Lam, William HK and Sumalee, Agachai and Li, Zhi-lin},
  journal={International Journal of Geographical Information Science},
  volume={26},
  number={2},
  pages={365--386},
  year={2012},
doi = {10.1080/13658816.2011.598133},
  publisher={Taylor \& Francis}
}

@article{kitsak2010identification,
  title={Identification of influential spreaders in complex networks},
  author={Kitsak, Maksim and Gallos, Lazaros K and Havlin, Shlomo and Liljeros, Fredrik and Muchnik, Lev and Stanley, H Eugene and Makse, Hern{\'a}n A},
  journal={Nature Physics},
  volume={6},
  number={11},
  pages={888--893},
  year={2010},
doi = {10.1038/nphys1746},
  publisher={Nature Publishing Group UK London}
}

@article{friedkin1991theoretical,
  title={Theoretical foundations for centrality measures},
  author={Friedkin, Noah E},
  journal={American Journal of Sociology},
  volume={96},
  number={6},
  pages={1478--1504},
  year={1991},
    doi = {10.1086/229694},
  publisher={University of Chicago Press}
}

@article{zhang2024mapreduce,
  title={A mapreduce-based approach for shortest path problem in road networks},
  author={Zhang, Dongbo and Shou, Yanfang and Xu, Jianmin},
  journal={Journal of Ambient Intelligence and Humanized Computing},
  pages={1--9},
  year={2024},
doi = {10.1007/s12652-018-0693-7},
  publisher={Springer}
}

@article{moukarzel1999spreading,
  title={Spreading and shortest paths in systems with sparse long-range connections},
  author={Moukarzel, Cristian F},
  journal={Physical Review E},
  volume={60},
  number={6},
  pages={R6263},
  year={1999},
doi = {10.1103/PhysRevE.60.R6263},
  publisher={APS}
}

@article{pei2013spreading,
  title={Spreading dynamics in complex networks},
  author={Pei, Sen and Makse, Hern{\'a}n A},
  journal={Journal of Statistical Mechanics: Theory and Experiment},
  volume={2013},
  number={12},
  pages={P12002},
  year={2013},
doi = {10.1088/1742-5468/2013/12/P12002},
  publisher={IOP Publishing}
}

@book{karunakaran2017spreading,
  title={Spreading Information in Complex Networks: An Overview and Some Modified Methods},
  author={Karunakaran, Reji Kumar and Manuel, Shibu and Satheesh, Edamana Narayanan},
  year={2017},
doi = {10.5772/intechopen.69204},
  publisher={IntechOpen}
}

@article{DU20133505,
title = {Effective usage of shortest paths promotes transportation efficiency on scale-free networks},
journal = {Physica A: Statistical Mechanics and its Applications},
volume = {392},
number = {17},
pages = {3505-3512},
year = {2013},
issn = {0378-4371},
doi = {10.1016/j.physa.2013.03.032},
author = {Wen-Bo Du and Zhi-Xi Wu and Kai-Quan Cai}
}

@article{VANVLIET19787,
title = {Improved shortest path algorithms for transport networks},
journal = {Transportation Research},
volume = {12},
number = {1},
pages = {7-20},
year = {1978},
issn = {0041-1647},
doi = {10.1016/0041-1647(78)90102-8},
author = {Dirck {Van Vliet}}
}

@incollection{pallottino1998shortest,
  title={Shortest path algorithms in transportation models: classical and innovative aspects},
  author={Pallottino, Stefano and Scutella, Maria Grazia},
  booktitle={Equilibrium and Advanced Transportation Modelling},
  pages={245--281},
  year={1998},doi={10.1007/978-1-4615-5757-9_11},
  publisher={Springer}
}

@article{johnson1977efficient,
  title={Efficient algorithms for shortest paths in sparse networks},
  author={Johnson, Donald B},
  journal={Journal of the ACM (JACM)},
  volume={24},
  number={1},
  pages={1--13},
  year={1977},doi = {10.1145/321992.321993},
  publisher={ACM New York, NY, USA}
}

@article{bellman1958routing,
  title={On a routing problem},
  author={Bellman, Richard},
  journal={Quarterly of Applied Mathematics},
  volume={16},
  number={1},
  pages={87--90},doi = {10.1090/qam/102435},
  year={1958}
}

@article{schrijver2012history,
  title={On the History of the Shortest Path Problem},
  author={Schrijver, Alexander},
  journal={Documenta Mathematica},
  volume={17},
  number={1},
  pages={155--167},doi = {10.4171/dms/6/19},
  year={2012}
}

@article{ford1956network,
  title={Network flow theory},
  author={Ford, Lester Randolph},
  journal={Rand Corporation Paper, Santa Monica, 1956},
  year={1956}
}

@article{FU20063324,
title = {Heuristic shortest path algorithms for transportation applications: State of the art},
journal = {Computers \& Operations Research},
volume = {33},
number = {11},
pages = {3324-3343},
year = {2006},
note = {Part Special Issue: Operations Research and Data Mining},
issn = {0305-0548},
doi = {10.1016/j.cor.2005.03.027},
author = {L. Fu and D. Sun and L.R. Rilett}
}

@article{wang1992analysis,
  title={Analysis of shortest-path routing algorithms in a dynamic network environment},
  author={Wang, Zheng and Crowcroft, Jon},
  journal={ACM SIGCOMM Computer Communication Review},
  volume={22},
  number={2},
  pages={63--71},
  year={1992},doi = {10.1145/141800.141805},
  publisher={ACM New York, NY, USA}
}

@article{magzhan2013review,
  title={A review and evaluations of shortest path algorithms},
  author={Magzhan, Kairanbay and Jani, Hajar Mat},
  journal={International Journal of Scientific and Technology Research},
  volume={2},
  number={6},
  pages={99--104},issn={2277-8616},
  year={2013}
}

@article{bertsekas1982dynamic,
  title={Dynamic behavior of shortest path routing algorithms for communication networks},
  author={Bertsekas, Dimitri},
  journal={IEEE Transactions on Automatic Control},
  volume={27},
  number={1},
  pages={60--74},
  year={1982},doi = {10.1109/TAC.1982.1102884},
  publisher={IEEE}
}

@ARTICLE{ChangAgeneticalgorithm,
  author={Chang Wook Ahn and Ramakrishna, R.S.},
  journal={IEEE Transactions on Evolutionary Computation}, 
  title={A genetic algorithm for shortest path routing problem and the sizing of populations}, 
  year={2002},
  volume={6},
  number={6},
  pages={566-579},
  doi={10.1109/TEVC.2002.804323}}

@book{penrose2003random,
  title={Random Geometric Graphs},
  author={Penrose, Mathew},
  volume={5},
  year={2003},
doi = {10.1093/acprof:oso/9780198506263.001.0001},
  publisher={Oxford University Press}
}

@INPROCEEDINGS{IslambekovROLE,
  author={Islambekov, Umar and Kumer Dey, Asim and Gel, Yulia R. and Poor, H. Vincent},
  booktitle={2018 IEEE Global Conference on Signal and Information Processing (GlobalSIP)}, 
  title={ROLE OF LOCAL GEOMETRY IN ROBUSTNESS OF POWER GRID NETWORKS}, 
  year={2018},
  volume={},
  number={},
  pages={885-889},
  doi={10.1109/GlobalSIP.2018.8646573}}

@Article{app8050772,
AUTHOR = {Ahmed, Saeed and Lee, YoungDoo and Hyun, Seung-Ho and Koo, Insoo},
TITLE = {Covert Cyber Assault Detection in Smart Grid Networks Utilizing Feature Selection and Euclidean Distance-Based Machine Learning},
JOURNAL = {Applied Sciences},
VOLUME = {8},
YEAR = {2018},
NUMBER = {5},
ARTICLE-NUMBER = {772},
ISSN = {2076-3417},
DOI = {10.3390/app8050772}
}

@INPROCEEDINGS{ZhangConnectivity,
  author={Zhang, Yu and Zhang, Hesheng and Sun, Wei and Pan, Cheng},
  booktitle={2014 IEEE Intelligent Vehicles Symposium Proceedings}, 
  title={Connectivity analysis for vehicular ad hoc network based on the Exponential Random Geometric Graphs}, 
  year={2014},
  volume={},
  number={},
  pages={993-998},
  doi={10.1109/IVS.2014.6856464}}

@article{buczkowska2019comparison,
  title={A comparison of euclidean distance, travel times, and network distances in location choice mixture models},
  author={Buczkowska, Sabina and Coulombel, Nicolas and de Lapparent, Matthieu},
  journal={Networks and Spatial Economics},
  volume={19},
  number={4},
  pages={1215--1248},
  year={2019},doi = {10.1007/s11067-018-9439-5},
  publisher={Springer}
}

@article{boyaci2021vehicle,
  title={Vehicle routing on road networks: How good is Euclidean approximation?},
  author={Boyac{\i}, Burak and Dang, Thu Huong and Letchford, Adam N},
  journal={Computers \& Operations Research},
  volume={129},
  pages={105197},
  year={2021},doi = {10.1016/j.cor.2020.105197},
  publisher={Elsevier}
}

@ARTICLE{VossDistributional,
  author={Voss, Florian and Gloaguen, Catherine and Fleischer, Frank and Schmidt, Volker},
  journal={IEEE Journal on Selected Areas in Communications}, 
  title={Distributional properties of euclidean distances in wireless networks involving road systems}, 
  year={2009},
  volume={27},
  number={7},
  pages={1047-1055},
  doi={10.1109/JSAC.2009.090903}}

@article{hekmat2006connectivity,
  title={Connectivity in wireless ad-hoc networks with a log-normal radio model},
  author={Hekmat, Ramin and Van Mieghem, Piet},
  journal={Mobile Networks and Applications},
  volume={11},
  pages={351--360},
  year={2006},doi = {10.1007/s11036-006-5188-7},
  publisher={Springer}
}

@article{friedrich2013diameter,
  title={Diameter and broadcast time of random geometric graphs in arbitrary dimensions},
  author={Friedrich, Tobias and Sauerwald, Thomas and Stauffer, Alexandre},
  journal={Algorithmica},
  volume={67},
  pages={65--88},
  year={2013},doi = {10.1007/s00453-012-9710-y},
  publisher={Springer}
}

@inproceedings{bradonjic2010efficient,
  title={Efficient broadcast on random geometric graphs},
  author={Bradonji{\'c}, Milan and Els{\"a}sser, Robert and Friedrich, Tobias and Sauerwald, Thomas and Stauffer, Alexandre},
  booktitle={Proceedings of the Twenty-first Annual ACM-SIAM Symposium on Discrete Algorithms},
  pages={1412--1421},
  year={2010},
doi={10.1137/1.9781611973075.114},
  organization={SIAM}
}

@article{ellis2007random,
  title={Random geometric graph diameter in the unit ball},
  author={Ellis, Robert B and Martin, Jeremy L and Yan, Catherine},
  journal={Algorithmica},
  volume={47},
  number={4},
  pages={421--438},
  year={2007},doi = {10.1007/s00453-006-0172-y},
  publisher={Springer}
}

@article{PenroseSRGG,
author = {Mathew D. Penrose},
title = {{Connectivity of soft random geometric graphs}},
volume = {26},
journal = {The Annals of Applied Probability},
number = {2},
publisher = {Institute of Mathematical Statistics},
pages = {986 -- 1028},
year = {2016},
doi = {10.1214/15-AAP1110}
}

@article{blasius2022efficiently,
  title={Efficiently generating geometric inhomogeneous and hyperbolic random graphs},
  author={Bl{\"a}sius, Thomas and Friedrich, Tobias and Katzmann, Maximilian and Meyer, Ulrich and Penschuck, Manuel and Weyand, Christopher},
  journal={Network Science},
  volume={10},
  number={4},
  pages={361--380},
  year={2022},doi = {10.1017/nws.2022.32},
  publisher={Cambridge University Press}
}

@article{bringmann2019geometric,
  title={Geometric inhomogeneous random graphs},
  author={Bringmann, Karl and Keusch, Ralph and Lengler, Johannes},
  journal={Theoretical Computer Science},
  volume={760},
  pages={35--54},
  year={2019},doi = {10.1016/j.tcs.2018.08.014},
  publisher={Elsevier}
}

@article{ErbaRandom,
  title = {Random geometric graphs in high dimension},
  author = {Erba, Vittorio and Ariosto, Sebastiano and Gherardi, Marco and Rotondo, Pietro},
  journal = {Physical Review E},
  volume = {102},
  issue = {1},
  pages = {012306},
  numpages = {7},
  year = {2020},
  month = {Jul},
  publisher = {American Physical Society},
  doi = {10.1103/PhysRevE.102.012306}
}

@article{dall2002random,
  title={Random geometric graphs},
  author={Dall, Jesper and Christensen, Michael},
  journal={Physical Review E},
  volume={66},
  number={1},
  pages={016121},
  year={2002},doi = {10.1103/PhysRevE.66.016121},
  publisher={APS}
}

@article{Kartun-GilesShape,
  title = {Shape of shortest paths in random spatial networks},
  author = {Kartun-Giles, Alexander P. and Barthelemy, Marc and Dettmann, Carl P.},
  journal = {Physical Review E},
  volume = {100},
  issue = {3},
  pages = {032315},
  numpages = {13},
  year = {2019},
  month = {Sep},
  publisher = {American Physical Society},
  doi = {10.1103/PhysRevE.100.032315}
}

@book{van2014performance,
  title={Performance Analysis of Complex Networks and Systems},
  author={Van Mieghem, P.},
  year={2014},
  publisher={Cambridge University Press},
doi={10.1017/CBO9781107415874},
ISBN={9781107415874},
  address={Cambridge, U.K.}
}

@article{dijkstra1959note,
  title={A note on two problems in connexion with graphs},
  author={Dijkstra, E. W.},
  journal={Numerische Mathematik},
  volume={1},
  number={1},
  pages={269--271},doi = {10.1007/BF01386390},
  year={1959}
}

@article{Raftopouloudronenetworks2022,
  title = {Fr\'echet distribution in geometric graphs for drone networks},
  author = {Raftopoulou, Maria and Litjens, Remco and Van Mieghem, Piet},
  journal = {Physical Review E},
  volume = {106},
  issue = {2},
  pages = {024301},
  numpages = {12},
  year = {2022},
  month = {Aug},
  publisher = {American Physical Society},
  doi = {10.1103/PhysRevE.106.024301}
}

@article{diaz2016relation,
author = {J. D{\'i}az and D. Mitsche and G. Perarnau and X. P{\'e}rez-Gim{\'e}nez},
title = {{On the relation between graph distance and Euclidean distance in random geometric graphs}},
volume = {48},
journal = {Advances in Applied Probability},
number = {3},
publisher = {Applied Probability Trust},
pages = {848 -- 864},doi = {10.1017/apr.2016.31},
year = {2016},
}

@book{cormen2022introduction,
  title={Introduction to Algorithms},
  author={Cormen, Thomas H and Leiserson, Charles E and Rivest, Ronald L and Stein, Clifford},
  year={2022},isbn={9780262046305},
  publisher={MIT press}
}

@article{kitsak2023finding,
  title={Finding shortest and nearly shortest path nodes in large substantially incomplete networks by hyperbolic mapping},
  author={Kitsak, Maksim and Ganin, Alexander and Elmokashfi, Ahmed and Cui, Hongzhu and Eisenberg, Daniel A and Alderson, David L and Korkin, Dmitry and Linkov, Igor},
  journal={Nature Communications},
  volume={14},
  number={1},
  pages={186},
  year={2023},doi = {10.1038/s41467-022-35181-w},
  publisher={Nature Publishing Group UK London}
}

@article{powers2020evaluation,
  title={Evaluation: from precision, recall and F-measure to ROC, informedness, markedness and correlation},
  author={Powers, David MW},
  journal={arXiv preprint arXiv:2010.16061},
  year={2020},
doi = {10.48550/arXiv.2010.16061
}
}

@article{vujovic2021classification,
  title={Classification model evaluation metrics},
  author={Vujovi{\'c}, {\v{Z}}},
  journal={International Journal of Advanced Computer Science and Applications},
  volume={12},
  number={6},
  pages={599--606},doi={10.14569/IJACSA.2021.0120670},
  year={2021}
}

@article{mao2012connectivity,
  title={Connectivity of large wireless networks under a general connection model},
  author={Mao, Guoqiang and Anderson, Brian DO},
  journal={IEEE Transactions on Information Theory},
  volume={59},
  number={3},
  pages={1761--1772},
  year={2012},doi = {10.1109/TIT.2012.2228894},
  publisher={IEEE}
}

@article{lu_vital_2016,
    title = {Vital nodes identification in complex networks},
    volume = {650},
    issn = {0370-1573},
    doi = {10.1016/j.physrep.2016.06.007},
    journal = {Physics Reports},
    author = {Lü, Linyuan and Chen, Duanbing and Ren, Xiao-Long and Zhang, Qian-Ming and Zhang, Yi-Cheng and Zhou, Tao},
    year = {2016},
    pages = {1--63},
}

@article{claffy2022challenges,
author = {kc Claffy and Clark, David},
doi = {10.5325/jinfopoli.12.2022.0003},
issn = {2381-5892},
journal = {Journal of Information Policy},
month = {jul},
pages = {195--233},
title = {{Challenges in Measuring the Internet for the Public Interest}},
volume = {12},
year = {2022}
}

@article{ahmed2020survey,
author = {Ahmed, Nadeem and Michelin, Regio A. and Xue, Wanli and Ruj, Sushmita and Malaney, Robert and Kanhere, Salil S. and Seneviratne, Aruna and Hu, Wen and Janicke, Helge and Jha, Sanjay K.},
doi = {10.1109/ACCESS.2020.3010226},
issn = {2169-3536},
journal = {IEEE Access},
pages = {134577--134601},
title = {{A Survey of COVID-19 Contact Tracing Apps}},
volume = {8},
year = {2020}
}

@article{luck2020reference,
author = {Luck, Katja and Kim, Dae-Kyum and Lambourne, Luke and Spirohn, Kerstin and Begg, Bridget E. and Bian, Wenting and Brignall, Ruth and Cafarelli, Tiziana and Campos-Laborie, Francisco J. and Charloteaux, Benoit and Choi, Dongsic and Cot{\'{e}}, Atina G. and Daley, Meaghan and Deimling, Steven and Desbuleux, Alice and Dricot, Am{\'{e}}lie and Gebbia, Marinella and Hardy, Madeleine F. and Kishore, Nishka and Knapp, Jennifer J. and Kov{\'{a}}cs, Istv{\'{a}}n A. and Lemmens, Irma and Mee, Miles W. and Mellor, Joseph C. and Pollis, Carl and Pons, Carles and Richardson, Aaron D. and Schlabach, Sadie and Teeking, Bridget and Yadav, Anupama and Babor, Mariana and Balcha, Dawit and Basha, Omer and Bowman-Colin, Christian and Chin, Suet-Feung and Choi, Soon Gang and Colabella, Claudia and Coppin, Georges and D'Amata, Cassandra and {De Ridder}, David and {De Rouck}, Steffi and Duran-Frigola, Miquel and Ennajdaoui, Hanane and Goebels, Florian and Goehring, Liana and Gopal, Anjali and Haddad, Ghazal and Hatchi, Elodie and Helmy, Mohamed and Jacob, Yves and Kassa, Yoseph and Landini, Serena and Li, Roujia and van Lieshout, Natascha and MacWilliams, Andrew and Markey, Dylan and Paulson, Joseph N. and Rangarajan, Sudharshan and Rasla, John and Rayhan, Ashyad and Rolland, Thomas and San-Miguel, Adriana and Shen, Yun and Sheykhkarimli, Dayag and Sheynkman, Gloria M. and Simonovsky, Eyal and Taşan, Murat and Tejeda, Alexander and Tropepe, Vincent and Twizere, Jean-Claude and Wang, Yang and Weatheritt, Robert J. and Weile, Jochen and Xia, Yu and Yang, Xinping and Yeger-Lotem, Esti and Zhong, Quan and Aloy, Patrick and Bader, Gary D. and {De Las Rivas}, Javier and Gaudet, Suzanne and Hao, Tong and Rak, Janusz and Tavernier, Jan and Hill, David E. and Vidal, Marc and Roth, Frederick P. and Calderwood, Michael A.},
doi = {10.1038/s41586-020-2188-x},
issn = {0028-0836},
journal = {Nature},
month = {apr},
number = {7803},
pages = {402--408},
pmid = {32296183},
title = {{A reference map of the human binary protein interactome}},
volume = {580},
year = {2020}
}

@article{Krioukov2010hyperbolic,
author = {Krioukov, Dmitri and Papadopoulos, Fragkiskos and Kitsak, Maksim and Vahdat, Amin and Bogu{\~{n}}{\'{a}}, Mari{\'{a}}n},
doi = {10.1103/PhysRevE.82.036106},
issn = {1539-3755},
journal = {Physical Review E},
month = {sep},
number = {3},
pages = {036106},
title = {{Hyperbolic geometry of complex networks}},
volume = {82},
year = {2010}
}

@article{Boguna2019small,
author = {Bogu{\~{n}}{\'{a}}, Mari{\'{a}}n and Krioukov, Dmitri and Almagro, Pedro and Serrano, M. {\'{A}}ngeles},
doi = {10.1103/PhysRevResearch.2.023040},
issn = {2643-1564},
journal = {Physical Review Research},
month = {apr},
number = {2},
pages = {023040},
title = {{Small worlds and clustering in spatial networks}},
volume = {2},
year = {2019}
}

@article{boguna_network_2021,
    title = {Network geometry},
    volume = {3},
    issn = {2522-5820},
    doi = {10.1038/s42254-020-00264-4},
    number = {2},
    journal = {Nature Reviews Physics},
    author = {Bogu{\~{n}}{\'{a}}, Mari{\'{a}}n and Bonamassa, Ivan and De Domenico, Manlio and Havlin, Shlomo and Krioukov, Dmitri and Serrano, M. {\'{A}}ngeles},
    month = feb,
    year = {2021},
    pages = {114--135},
}

@article{van2001paths,
  title={Paths in the simple random graph and the Waxman graph},
  author={Van Mieghem, Piet},
  journal={Probability in the Engineering and Informational Sciences},
  volume={15},
  number={4},
  pages={535--555},
  year={2001},doi = {10.1017/S0269964801154070},
  publisher={Cambridge University Press}
}

@article{hooghiemstra2005mean,
  title={On the mean distance in scale free graphs},
  author={Hooghiemstra, G and Van Mieghem, P},
  journal={Methodology and Computing in Applied Probability},
  volume={7},
  number={3},
  pages={285--306},
  year={2005},doi = {10.1007/s11009-005-4518-8},
  publisher={Springer}
}

@article{voitalov2020weighted,
  title={Weighted hypersoft configuration model},
  author={Voitalov, Ivan and Van Der Hoorn, Pim and Kitsak, Maksim and Papadopoulos, Fragkiskos and Krioukov, Dmitri},
  journal={Physical Review Research},
  volume={2},
  number={4},
  pages={043157},
  year={2020},
doi = {10.1103/PhysRevResearch.2.043157},
  publisher={APS}
}

@article{allard2017geometric,
  title={The geometric nature of weights in real complex networks},
  author={Allard, Antoine and Serrano, M {\'A}ngeles and Garc{\'\i}a-P{\'e}rez, Guillermo and Bogu{\~n}{\'a}, Mari{\'a}n},
  journal={Nature Communications},
  volume={8},
  number={1},
  pages={14103},
  year={2017},
doi = {10.1038/ncomms14103},
  publisher={Nature Publishing Group UK London}
}

@article{simini2012universal,
  title={A universal model for mobility and migration patterns},
  author={Simini, Filippo and Gonz{\'a}lez, Marta C and Maritan, Amos and Barab{\'a}si, Albert-L{\'a}szl{\'o}},
  journal={Nature},
  volume={484},
  number={7392},
  pages={96--100},
  year={2012},
doi = {10.1038/nature10856},
  publisher={Nature Publishing Group UK London}
}

\end{document}